\documentclass[aps,prl,preprint,superscriptaddress]{revtex4-1}
\usepackage{amssymb}
\usepackage[utf8]{inputenc}
\usepackage{graphicx}
\usepackage{bm,color,subfigure,amsmath}

\begin{document}

\title{Reconfigurable spin wave band structure of artificial square spin ice}

\author{Ezio~Iacocca}
\email{ezio.iacocca@colorado.edu}
\affiliation{Department of Applied Mathematics, University of Colorado, Boulder, Colorado 80309-0526, USA}
\affiliation{Department of Applied Physics, Division for condensed matter theory, Chalmers University of Technology, 412 96, Gothenburg, Sweden}
\affiliation{Physics Department, University of Gothenburg, 412 96, Gothenburg, Sweden}

\author{Sebastian~Gliga}
\affiliation{Laboratory for Mesoscopic Systems, Department of Materials, ETH Zurich, 8093 Zurich, Switzerland}
\affiliation{Laboratory for Micro- and Nanotechnology, Paul Scherrer Institute, 5232 Villigen PSI, Switzerland}

\author{Robert~L.~Stamps}
\affiliation{School of Physics and Astronomy, University of Glasgow, Glasgow, UK, G12 8QQ}

\author{Olle~Heinonen}
\affiliation{Materials Science Division, Argonne National Laboratory, Lemont, IL 60439, USA}
\affiliation{Northwestern-Argonne Institute for Science and Engineering, Evanston, IL 60208, USA}

\begin{abstract}
Artificial square spin ices are structures composed of magnetic elements arranged on a geometrically frustrated lattice and located on the sites of a two-dimensional square lattice, such that there are four interacting magnetic elements at each vertex. Using a semi-analytical approach, we show that square spin ices exhibit a rich spin wave band structure that is tunable both by external magnetic fields and the configuration of individual elements. Internal degrees of freedom can give rise to equilibrium states with bent magnetization at the edges leading to characteristic excitations; in the presence of magnetostatic interactions these form separate bands analogous to impurity bands in semiconductors. Full-scale micromagnetic simulations corroborate our semi-analytical approach. Our results show that artificial square spin ices can be viewed as reconfigurable and tunable magnonic crystals that can be used as metamaterials for spin-wave-based applications at the nanoscale.
\end{abstract}
\maketitle

Spin waves, or magnons, are fundamental excitations in magnetic thin films and nanostructures. Because of their potential applications in information technology~\cite{Kostylev2005,Khitun2005} and computation~\cite{Chumak2013}, means to control magnon dispersion and band gap have been studied intensively over the past few decades. The term {\em magnonics} has been coined to describe this field of study~\cite{Demokritov2013,Krawczyk2014}. One pathway to control magnon dispersions is to construct magnonic crystals~\cite{Nikitov2001,Neusser2009} that are metamaterials with a spatial modulation of the magnetic properties on length scales comparable to relevant magnonic wavelengths~\cite{Wang2010,Tacchi2011,Tacchi2012}. Patterned thin magnetic films~\cite{Kruglyak2010,Lenk2011} or topographically modulated thin films have been used to manipulate the magnon spectra~\cite{Sklenar2015}. This approach is similar to super-lattices in photonics and, fundamentally, to the crystal structure of semiconductors. A paradigm that is the focus of recent investigation consists in actively modifying the band structure of magnonic crystals~\cite{Grundler2015}. This has been achieved to date by use of Meander-type structures~\cite{Karenowska2012} and, more recently, via heating~\cite{Vogel2010} in one-dimensional ferromagnets.

\begin{figure}[t] 
\centering \includegraphics[width=3.in]{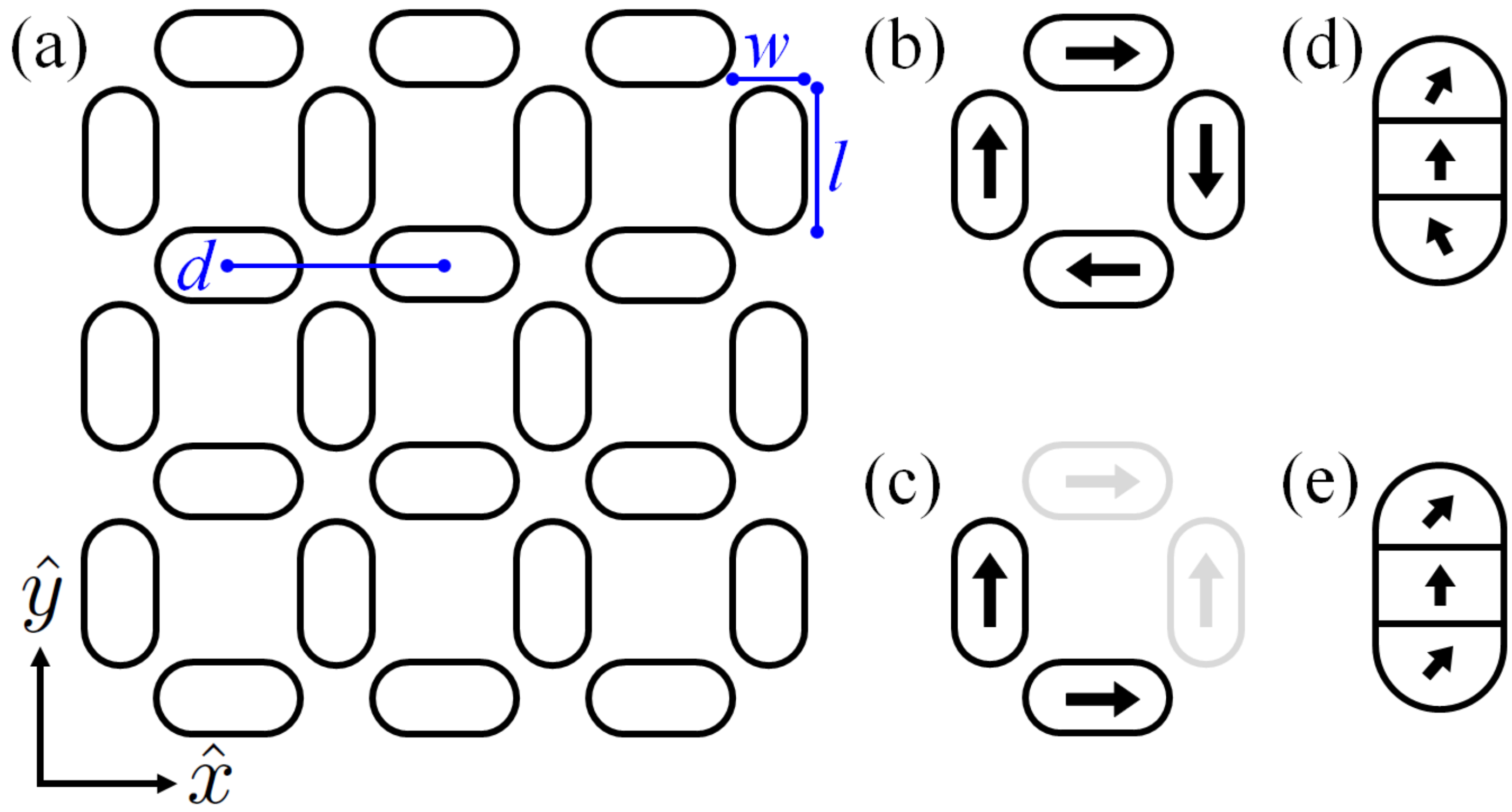}
\caption{ \label{fig1} (Color online) (a) Square ice lattice with lattice constant $d$, and with magnetic stadia of width $w$, length $l$, and thickness $t$. The stable magnetization directions (black arrows) of the magnetic elements in a unit cell are shown for the (b) ground (vortex) and (c) remanent state (the gray-colored stadia are not part of the unit cell and are shown here for clarity). Edge states have two stable configurations as (d) C and (e) S states. }
\end{figure}

Artificial spin ices~\cite{Wang2005,Nisoli2013,Heyderman2013} are another class of structures based on an organized array of nanosized magnetic elements that have been shown to support a wealth of static, dynamic, and emergent magnetic phenomena~\cite{Heyderman2013,Roadmap2014,Kapaklis2014}. Artificial spin ices are geometrically {\em frustrated}: the geometry of the elements and the lattice are such that all interaction energies cannot be simultaneously minimized. Examples of artificial spin ices are the square ice~\cite{Wang2005}, and the kagome ice~\cite{Qi2008}. The square ice is composed of magnetic stadium-shaped nanoislands positioned on the sites of a two-dimensional square lattice with lattice constant $d$, Fig.~\ref{fig1}(a), and obeys the ``ice rules'' in which low-energy states are characterized by the magnetization in two islands pointing into a vertex and out of the vertex in the two other nanoislands. Dynamically, correlated excitations are supported in spin ices because of the magnetostatic interactions between magnetic islands~\cite{Gliga2013}. Because of their intrinsic periodicity and wealth of static states, artificial spin ices offer interesting opportunities as programmable magnonic crystals to control the magnon dispersion and band gap ~\cite{Heyderman2013}.

The resonant mode spectrum of square ices has been studied numerically by means of micromagnetic simulations, demonstrating the observable effects of magnetic defects~\cite{Gliga2013}. More recently, a detailed numerical study has shown that edge modes arising from the internal degrees of freedom equally have observable consequences in the resonant spectrum in sufficiently thick nanoislands~\cite{Gliga2015}. In fact, edge modes efficiently couple neighboring nanoislands, influencing the collective oscillations~\cite{Heyderman2013}. This is reminiscent of impurity states in semiconductors that locally modify the energy landscape and give rise to shallow electronic bands~\cite{Balkanski2000}. Recent experimental results have explored the excitation spectrum of artificial spin ices~\cite{Jungfleisch2016,Bhat2016,Zhou2016} but the existence and dependencies of the band diagram in square ices has not been explored to date. To close the gap between the fields magnonics and artificial spin ices, we examine square ices from the perspective of magnonics, including bands arising from the edge modes as well as the bulk modes.

In this Rapid Communication, we study long-range dipolar-mediated two-dimensional magnon dispersion in square ices in the spirit of a tight-binding model. In contrast to similar procedures on simpler structures~\cite{Shindou2013,Shindou2013b}, we account for the internal degrees of freedom resulting from edge modes in the nanoislands. Consequently, we are able to calculate the magnon dispersion as a function of local equilibrium states as well as its field tunability, including edge mode bands. Our semi-analytical approach provides enough degrees of freedom to qualitatively estimate the band structure of an extended square ice lattice while being computationally tractable.

We focus on the small-amplitude excitations in two energetically stable configurations of a square ice, namely the vortex and remanent states, Fig.~\ref{fig1}(b) and (c), respectively. The vortex state is the ground state of the system, achieved by thermal relaxation~\cite{Farhan2013}, and the remanent state can be obtained by saturating the system in an external field along the $(\hat{x},\hat{y})$ direction, and then slowly removing the external field, letting the system relax. In each configuration, the magnetization can bend close to the nanoisland edges~\cite{Gliga2015}, providing a local ``impurity''. In square ices, two stable edge configurations satisfy the minimization of dipolar fields at the ground state, resulting in C and S states~\cite{Cowburn2000,Madami2011b,Carlotti2014}, Fig.~\ref{fig1}(d)-(e).

The small-amplitude dynamics in square ices can be approached semi-analytically using a Hamiltonian formalism~\cite{Slavin2009}. The same approach has been used and shown to be accurate in many dynamical regimes to date~\cite{Slavin2005,Bonetti2010,Iacocca2012,Iacocca2013b,Iacocca2014b,Iacocca2015,Locatelli2015}. In this formalism, the Landau-Lifshitz equation of motion describing conservative magnetization dynamics is cast as a function of a complex amplitude $a$, using a Holstein-Primakoff transformation.
By expanding the resulting equation in Taylor series, the linear dynamics for an ensemble of complex amplitudes $\underline{a}$ can be generally expressed (see Supplementary material) as
\begin{equation}
\label{eq:linear}
    \frac{d\underline{a}}{dt} = -i\frac{d}{d\underline{a}^*}\underline{A}^\dagger\mathcal{H}\underline{A} = -i\frac{d}{d\underline{a}^*}\underline{A}^\dagger\begin{pmatrix}
                                                    \mathcal{H}^{(1,1)} & \mathcal{H}^{(1,2)} \\
                                                    \mathcal{H}^{(2,1)} & \mathcal{H}^{(2,2)} \\
                                                 \end{pmatrix}\underline{A},
\end{equation}
where the dagger denotes the complex transpose, $\underline{A}$ is an array of $2n$ complex amplitudes $\underline{A}^T=[\underline{a}^T,\underline{a}^\dagger]=[a_1, ..., a_n, a_1^*, ..., a_n^*]$ and $\mathcal{H}$ is the $2n\times 2n$ Hamiltonian. The right-hand-side of Eq.~(\ref{eq:linear}) includes terms up to second order in $\underline{a}$, corresponding to linear excitations. Beacuse of the lattice perodicity, propagating waves are Bloch waves with a time dependence gvien by $a\rightarrow ae^{i\omega t}$. This allows us to reduce Eq.~(\ref{eq:linear}) to an eigenvalue problem by means of Colpa's grand dynamical matrix~\cite{Colpa1978}
\begin{equation}
\label{eq:eigen}
    \omega\underline{\psi} = \begin{pmatrix}
                            \mathcal{H}^{(1,2)} & \mathcal{H}^{(2,2)} \\
                            \mathcal{H}^{(1,1)} & \mathcal{H}^{(2,1)} \\
                          \end{pmatrix}\underline{\psi},
\end{equation}
from which we obtain the eigenvalues $\omega$, and the eigenvectors $\underline{\psi}$. Due to the complex conjugate definition of $\underline{A}$, we observe that $\mathcal{H}^{(1,1)}=\mathcal{H}^{(2,2)}$ and $\mathcal{H}^{(1,2)}=\mathcal{H}^{(2,1)}$, leading to conjugate eigenvalues in Eq.~(\ref{eq:eigen}).

The Hamiltonian is related to the magnetic field $\vec{H}$ via $\mathcal{H}=-\gamma\delta W/(2M_S)$, where $\delta W=-\int \vec{H}(\vec{M}) \cdot d\vec{M}$ is the energy functional, $\gamma$ is the gyromagnetic ratio, $\vec{M}$ is the magnetization vector, and $M_S=||\vec{M}||$ is the saturation magnetization. We consider field contributions from shape anisotropy, dipolar interactions, and intra-element exchange as well as an external applied field. Each field contribution can be reduced to a Hamiltonian matrix as detailed in the Supplementary material. Of particular importance are the dipolar interactions, which are the only source of inter-element coupling in our framework and the concomitant magnon dispersion. The dipolar energy between a nanoisland $j$ in cell $\tau$ and all the other nanoislands $k$ in cells $\tau'$ can be expressed as
\begin{eqnarray}
\label{eq:dipolar}
  \mathcal{H}_d &=& -\frac{V}{4\pi}\sum_{k,\tau'}\Big[\frac{3(\vec{R}_{jk,\tau\tau'}\cdot\vec{M}_{j,\tau'})(\vec{R}_{jk,\tau\tau'}\cdot\vec{M}_{j,\tau})}{(\vec{R}_{jk,\tau\tau'})^5}\nonumber\\&-&\frac{\vec{M}_{j,\tau'}\cdot\vec{M}_{j,\tau}}{(\vec{R}_{jk,\tau\tau'})^3}\Big],
\end{eqnarray}
where $V$ is the volume of the magnetic element and $\vec{R}_{jk,\tau\tau'}$ is the translation vector between the nanoisland $j$ in cell $\tau$ and the nanoisland $k$ in cell $\tau'$. Considering the Bloch wave $\vec{M}_{j,\tau}=\vec{M}_{j,\tau'}e^{i\vec{q}\vec{R}_{jk}}$, where $\vec{q}$ is the wave vector, it is possible to recast Eq.~(\ref{eq:dipolar}) for the unit cell in terms of the lattice ($S_\beta$ where $\beta = \hat{x},\hat{y},\hat{z}$) and cross-direction ($S_c$) summations. As an example, the resulting Hamiltonian matrices for the ground state in the absence of exchange interactions are
\begin{subequations}
\label{eq:Hd}
\begin{eqnarray}
\label{eq:Hd11}
  \mathcal{H}_{d}^{(1,1)} &=&   \begin{pmatrix}
	                                S_x^{11} & S_c^{12} & S_x^{13} & S_c^{14} \\
									S_c^{21} & S_y^{22} & S_c^{23} & S_y^{24} \\
									S_x^{31} & S_c^{23} & S_x^{33} & S_c^{34} \\
									S_c^{41} & S_y^{24} & S_c^{34} & S_y^{44} \\
	                            \end{pmatrix} - \hat{S}_z\\
\label{eq:Hd21}
  \mathcal{H}_{d}^{(1,2)} &=&   \mathcal{D}+\begin{pmatrix}
	                                        0 & S_c^{12} & S_x^{13} & S_c^{14} \\
											S_c^{21} & 0 & S_c^{23} & S_y^{24} \\
											S_x^{31} & S_c^{32} & 0 & S_c^{34} \\
											S_c^{41} & S_y^{42} & S_c^{43} & 0 \\
	                                        \end{pmatrix}+ \hat{S}_z,
\end{eqnarray}
\end{subequations}
where $\mathcal{D}$ is a diagonal matrix containing inter-island interactions (the expressions for $\mathcal{D}$ and the lattice summations are shown in the Supplementary material). The reduction of the dipolar field to Hamiltonian matrices is a key result of this work.

The magnon dispersion can be numerically calculated by solving the eigenvalue problem of Eq.~(\ref{eq:eigen}). We consider a square ice composed of Permalloy stadia with dimensions $280$~nm$\times$~$120$~nm$\times$~$20$~nm, saturation magnetization $M_S=770$~kA/m, and center-to-center separation of $d=395$~nm. Exchange is implemented as an additional degree of freedom in a nanoisland divided by three equidistant spins coupled by the constant $J=0.016$, which parametrizes the exchange in Permalloy $J=cA/2$, where $c=0.33$~nm is the lattice constant and $A=10$~pJ/m is the exchange stiffness. This approximation for the exchange interaction is applicable for the low-energy sector of the magnon bands, as demonstrated below by the good quantitative agreement with full-scale micromagnetic simulations.
\begin{figure}[t] 
\centering \includegraphics[width=3.in]{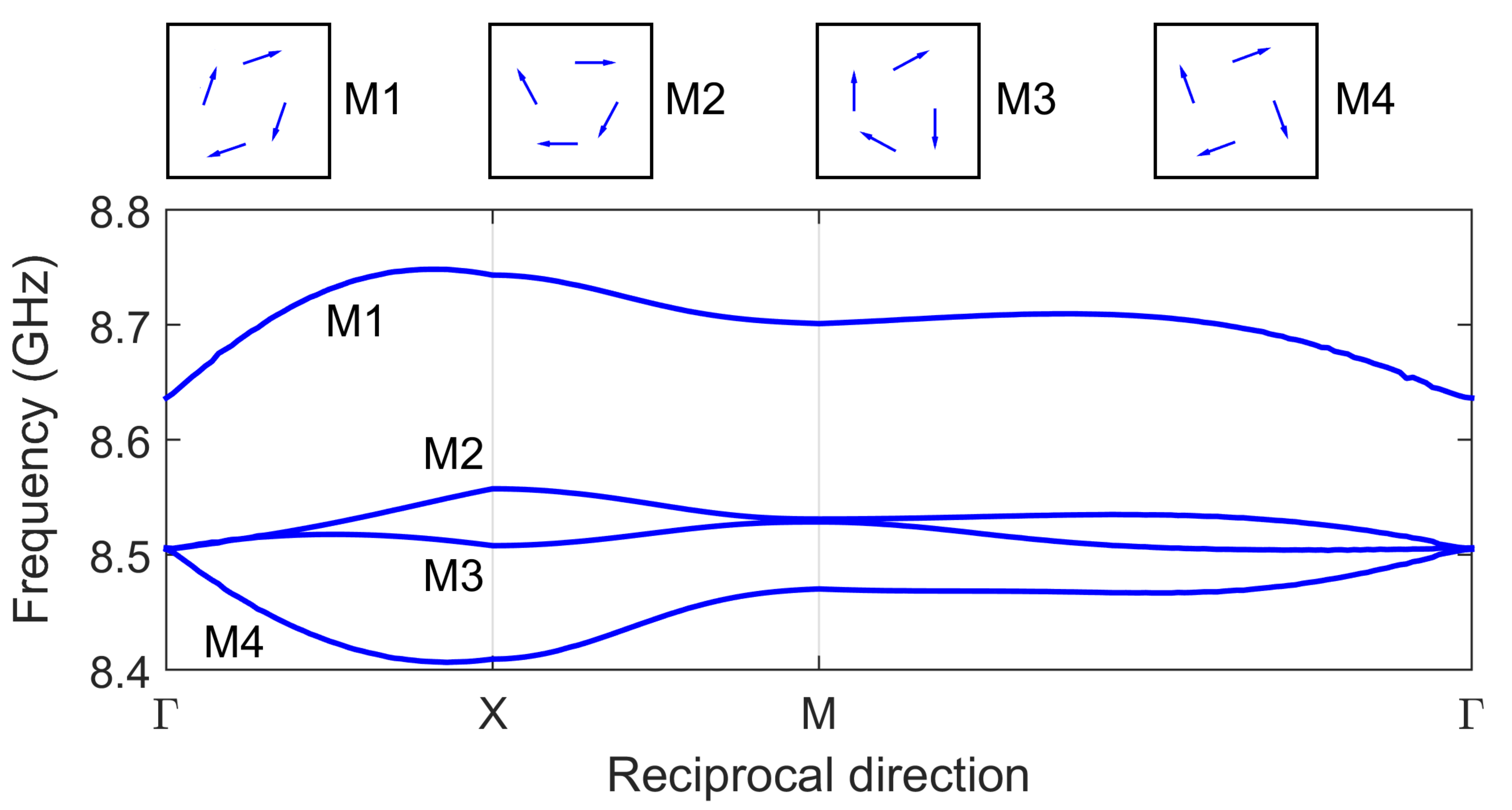}
\caption{ \label{fig2} (Color online) Band structure of the vortex state. The insets show the magnetization vector configuration of the unit cell for each band, showcasing their excitation symmetry. }
\end{figure}

It is instructive to consider first the band structure neglecting internal degrees of freedom, or ``macrospin'' approximation. A typical band structure for the macrospin vortex state is shown in Fig.~\ref{fig2}. There are four bands consistent with the available degrees of freedom in the system, one for each island. From the corresponding eigenvectors, it is possible to identify the location and symmetry of each mode. A snapshot of the magnetic configurations at the $\Gamma$ point for each band (labeled from M1 to M4) are shown above Fig.~\ref{fig2}. We notice that M1 has pair of islands in phase and a phase difference of $\pm\pi$ between each pair, whereas M4 represents a mode with all islands excited in phase. Furthermore, M1 (M4) has positive (negative) group velocity. M2 and M3 are close in energy and consist of modes with a pairwise phase difference of $\pm\pi/2$. Note that the pairwise difference make these bands non-degenerate, resulting in anti-crossings close to the $\Gamma$ and M points. These latter two modes form narrow bands that separate away form the $\Gamma$ point, and establish a band gap reaching $\approx 195$~MHz between the $\Gamma$ and X point of M1 and M2, respectively. Bands effectively touch at the $\Gamma$ and $M$ points. However, we did not observe band inversion in any calculation.

\begin{figure}[t] 
\centering \includegraphics[width=3.in]{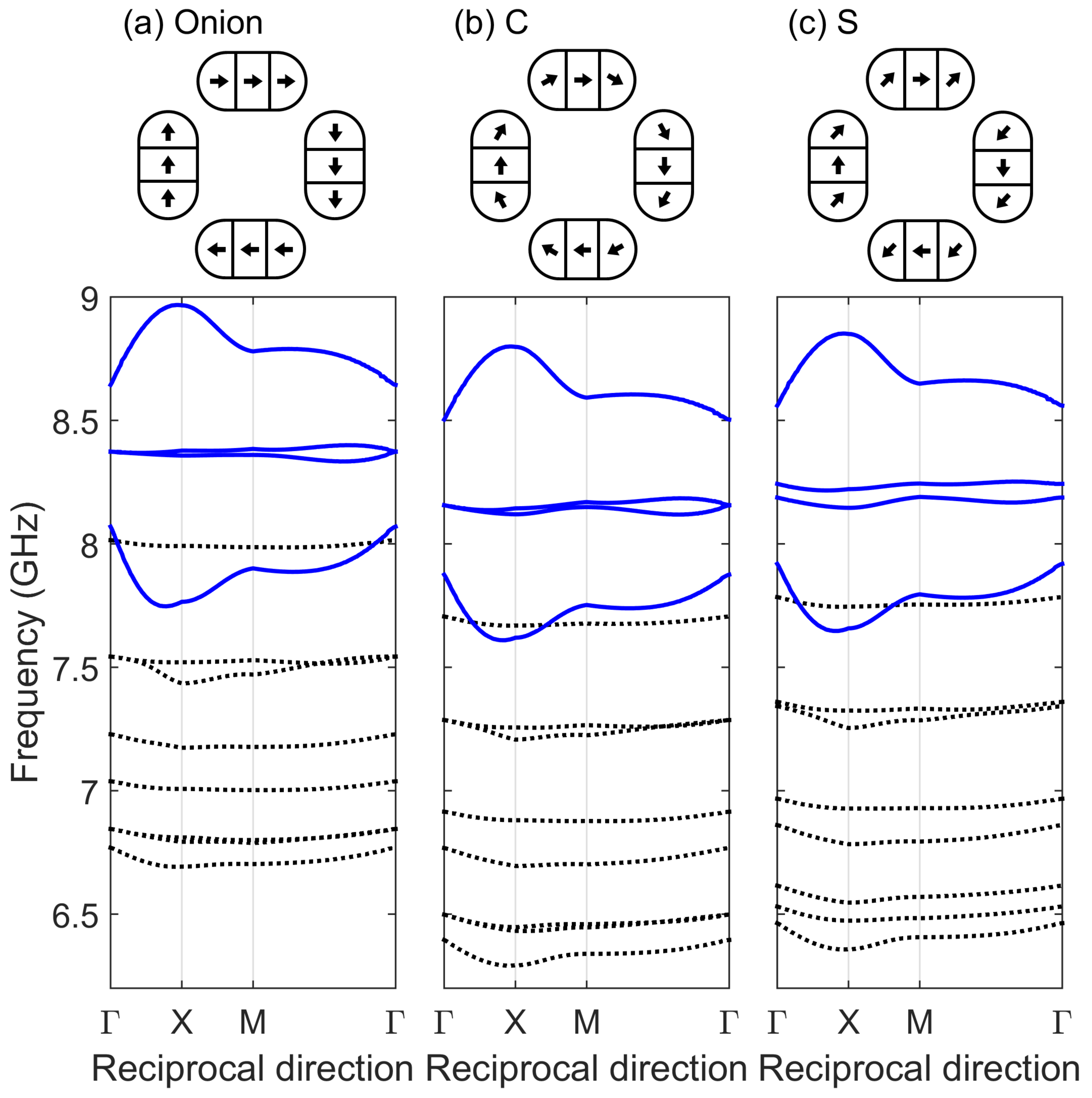}
\caption{ \label{fig3} (Color online) Band diagram for the vortex state in a (a) onion, (b) C, and (c) S state. The bulk (edge) modes are depicted in blue (dashed black) lines. The schematic of each static configuration is also shown for each case. }
\end{figure}

\begin{figure}[t] 
\centering \includegraphics[width=3.in]{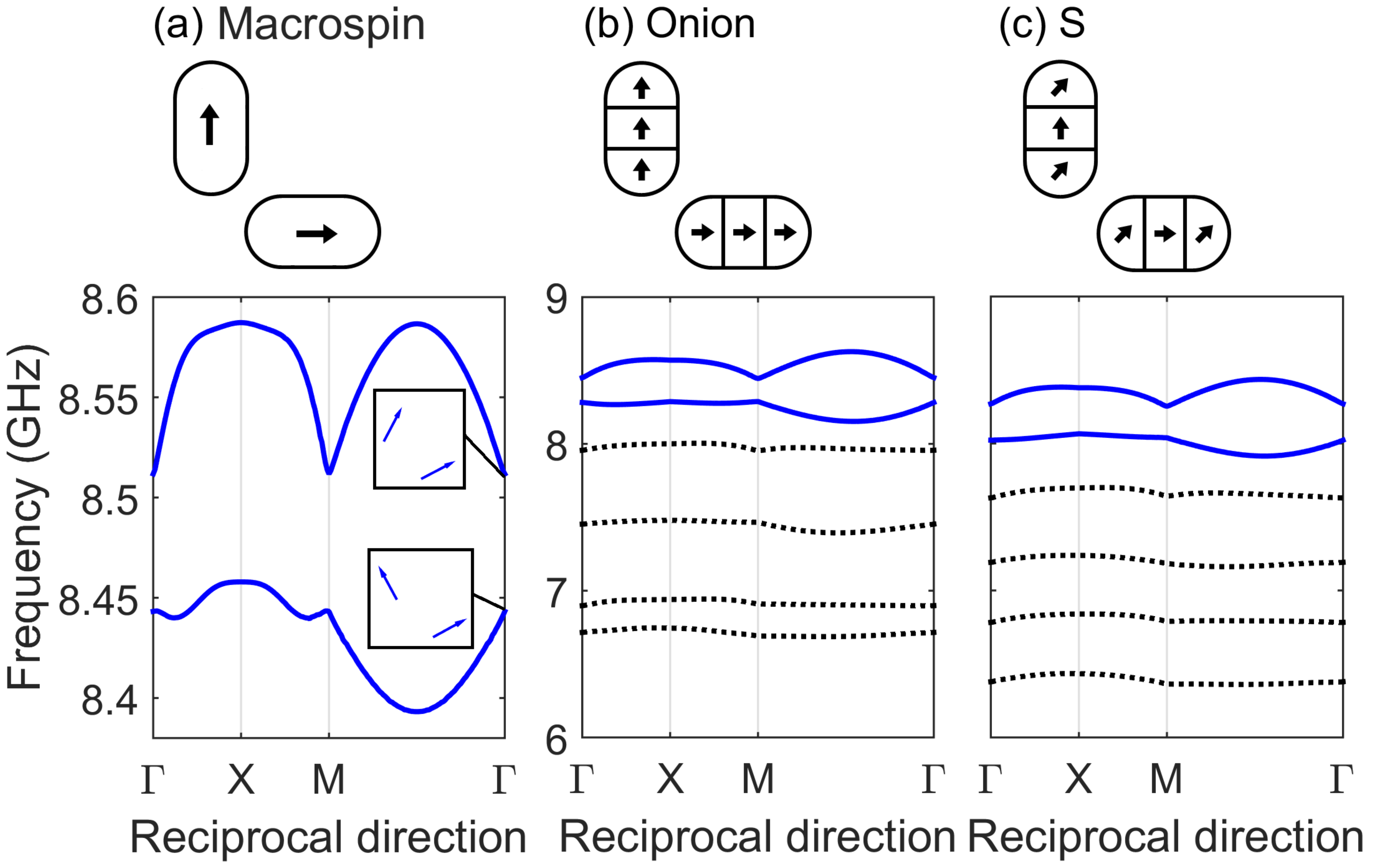}
\caption{ \label{fig4} (Color online) Band diagram for the remanent state in a (a) macrospin, (b) onion, and (c) S states. The bulk (edge) modes are depicted in blue (dashed black) lines. The inset shows the magnetization vector configuration of the unit cell for each band. The schematic of each static configuration is also shown for each case. }
\end{figure}

We now include exchange interactions in our framework. By dividing each magnetic island into three equidistant spins, we now have access to 12 bands. In the ground state, three configurations are stable: homogeneous or onion~\cite{Gliga2015}, C, and S states. The corresponding band diagrams are shown in Fig.~\ref{fig3}. The additional degrees of freedom give rise to lower frequency bands identified as edge modes (black dashed lines), also showing anti-crossing behavior. We observe that the bulk modes (blue lines) maintain their qualitative features. However, the bandgaps are enhanced due to the additional energy incorporated into the system. Furthermore, the particular magnetic configuration quantitatively modifies the band diagram, indicating that edge bending can be compared to impurity states in semiconductor materials. Because a transition between C and S states can be induced by, {\em e.g.,} temperature~\cite{Gliga2015}, this can be used as another avenue to program the magnonic response of the square ice. In the remanent state, the unit cell is composed of two magnetic islands, Fig.~\ref{fig1}(c). The band diagrams for a macrospin and stable onion and S configurations are shown in Fig.~\ref{fig4}, exhibiting similar features as discussed above.

\begin{figure}[t] 
\centering \includegraphics[width=3.in]{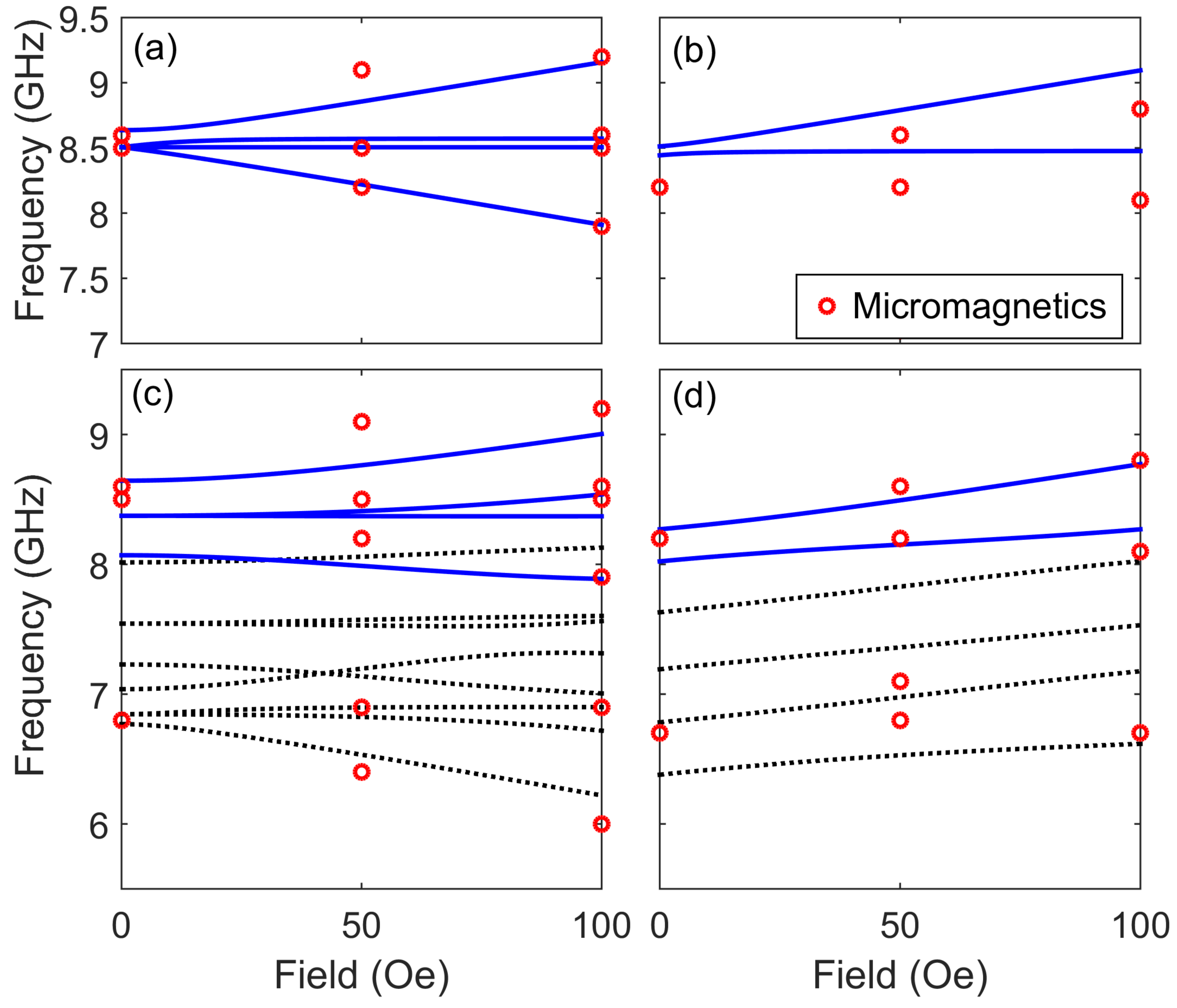}
\caption{ \label{fig5} (Color online) Magnon frequencies at the $\Gamma$ point as a function of an external field applied along the $\hat{x}$ axis for the (a) macrospin vortex, (b) macrospin remanent, (c) onion vortex, and (d) remanent S states. The bulk (edge) modes are depicted in blue (dashed black) lines. The red dots are obtained from micromagnetic simulations. }
\end{figure}

We now explore the effect of an applied field $\vec{H}_e$ on the square ice. We consider a feasible experimental scenario of an in-plane field along the $\hat{x}$ direction and detection of coherent excitations (at the $\Gamma$ point) by means of resonance measurements (The effect of field angle is shown in the Supplementary material). Note that in our framework, the stable magnetization direction of the magnetic nanoislands is set and assumed {\it a priori}, \emph{i.e.}, only small amplitude variations are accessible. In fact, large fields induce imaginary eigenvalues, denoting decaying modes and thus the breakdown of our model. We study the effect of field magnitudes between $0<|\vec{H}_e|<100$~Oe which maintains real eigenvalues. The results obtained for both vortex and remanent states under macrospin approximation are shown in Fig.~\ref{fig5}(a-b). In the case of the vortex state, we observe that the coherent modes, M1 and M4, have positive and negative tunabilities, respectively, whereas M2 and M3 exhibit only slight tunability. In the case of the remanent state we observe either a positive or negligible tunability. The strongly tunable modes can be traced to those magnetic elements parallel to the applied field. This is also consistent with the Landau-Lifshitz equation predicting a blue (red) shift of frequencies when the \emph{internal} field increases (decreases). The modes with negligible tunability correspond to magnetic elements perpendicular to the field. By considering edge bending, a richer behavior for the tunability of both the vortex and remanent states is obtained, Fig.~\ref{fig5}(c-d). For both the vortex in an onion state and the remanent S state, we observe similar tunabilities for the bulk and low-frequency edge modes. In all cases, the slope of each band is generally different, leading to band crossings, and implying that the bandgaps in square ices can be manipulated by an applied magnetic field.

Full-scale micromagnetic simulations were performed for comparison with the semi-analytical model. We used a computational system containing eight islands and imposing periodic boundary conditions consistent with the geometry described above (see Supplemental Material for details). The results are shown as red circles in Fig.~\ref{fig5} (note that the micromagnetic modeling only returns modes that are even in the unit cell because the exciting field is uniform, while the semi-analytical model captures all modes irrespective of symmetry). For the vortex state, a good agreement for the bulk modes is obtained from the macrospin model. Further comparison with the extended semi-analytical model also shows excellent agreement with the low-frequency edge modes. For the remanent state, the macrospin model yields a good qualitative agreement with the micromagnetic results. A three-spin S-state model also yields good agreement with the micromagnetic low-frequency modes, especially in view of the simplistic treatment in the three-spin model of the smooth static equilibrium magnetization in the micromagnetic model.

We remark that in both real and micromagnetically-modeled nanoislands, there are many higher-order modes, beyond what can be described by the three-spin model considered here, because of the many internal degrees of freedom. Such higher-order modes have many internal nodal lines of the magnetization eigenmodes. Therefore, the magnetostatic fields emanating from such modes decay rather quickly in space. This results in a weak coupling between different islands so the magnonic bands arising from such modes are non-dispersive with no phase or group velocity, and are not interesting here. It is also noteworthy that a strong variation of the islands aspect ratio can significantly affect the excited frequencies, {\em i.e.}, in the nanowire and circular dot limits. Moreover, we expect the thickness to play an important role in the ultra-thin film regime, where the anisotropy becomes perpendicular or can favor a vortex state inside each stadium.

In summary, we have calculated the magnon band structure and the mode tunability at the $\Gamma$ point for a square ice in two equilibrium states, the vortex (or ground) state, and the remanent state, using a model that includes internal degrees of freedom of the islands as well as edge bending. The good quantitative agreement with micromagnetic simulations confirms the accuracy of the small-amplitude semi-analytical model while avoiding the computational limitations intrinsic to fully three-dimensional micromagnetic simulations. These results show that the magnon spectra, and therefore group and phase velocities as well as band gap, can be manipulated by external fields. In particular, the edge modes give rise to separate magnon bands allowing for a larger parameter space in terms of magnon control. This suggests that square ices can be considered metamaterials for spin waves. In addition, the square ice is in principle reconfigurable in that the magnetization in individual islands can be changed by the application of external fields ({\em e.g.,} from vortex to remanent state) or temperature, or by using more sophisticated techniques such as using spin torque by patterning nanocontacts on the elements or making the elements part of magnetic tunnel junctions. This opens up the possibility of two-dimensional reprogrammable magnonic crystals comprised of an artificial square spin ice.

\begin{acknowledgments}
E.~I. acknowledges support from the Swedish Research Council, Reg. No. 637-2014-6863. The work by O.~H. was funded by the Department of Energy Office of Science, Materials Sciences and Engineering Division. E.~I. was partly supported by the U.S. Department of Energy, Office of Science, Materials Sciences and Engineering Division through the Materials Theory Institute. We gratefully acknowledge the computing resources provided on Blues, a high-performance computing cluster operated by the Laboratory Computing Resource Center at Argonne National Laboratory. The work by R.~L.~S. was funded by EPSRC EP/L002922/1.
\end{acknowledgments}

\bibliographystyle{aipnum4-1}

\begin{thebibliography}{43}%
\makeatletter
\providecommand \@ifxundefined [1]{%
 \@ifx{#1\undefined}
}%
\providecommand \@ifnum [1]{%
 \ifnum #1\expandafter \@firstoftwo
 \else \expandafter \@secondoftwo
 \fi
}%
\providecommand \@ifx [1]{%
 \ifx #1\expandafter \@firstoftwo
 \else \expandafter \@secondoftwo
 \fi
}%
\providecommand \natexlab [1]{#1}%
\providecommand \enquote  [1]{``#1''}%
\providecommand \bibnamefont  [1]{#1}%
\providecommand \bibfnamefont [1]{#1}%
\providecommand \citenamefont [1]{#1}%
\providecommand \href@noop [0]{\@secondoftwo}%
\providecommand \href [0]{\begingroup \@sanitize@url \@href}%
\providecommand \@href[1]{\@@startlink{#1}\@@href}%
\providecommand \@@href[1]{\endgroup#1\@@endlink}%
\providecommand \@sanitize@url [0]{\catcode `\\12\catcode `\$12\catcode
  `\&12\catcode `\#12\catcode `\^12\catcode `\_12\catcode `\%12\relax}%
\providecommand \@@startlink[1]{}%
\providecommand \@@endlink[0]{}%
\providecommand \url  [0]{\begingroup\@sanitize@url \@url }%
\providecommand \@url [1]{\endgroup\@href {#1}{\urlprefix }}%
\providecommand \urlprefix  [0]{URL }%
\providecommand \Eprint [0]{\href }%
\providecommand \doibase [0]{http://dx.doi.org/}%
\providecommand \selectlanguage [0]{\@gobble}%
\providecommand \bibinfo  [0]{\@secondoftwo}%
\providecommand \bibfield  [0]{\@secondoftwo}%
\providecommand \translation [1]{[#1]}%
\providecommand \BibitemOpen [0]{}%
\providecommand \bibitemStop [0]{}%
\providecommand \bibitemNoStop [0]{.\EOS\space}%
\providecommand \EOS [0]{\spacefactor3000\relax}%
\providecommand \BibitemShut  [1]{\csname bibitem#1\endcsname}%
\let\auto@bib@innerbib\@empty
\bibitem [{\citenamefont {Kostylev}\ \emph {et~al.}(2005)\citenamefont
  {Kostylev}, \citenamefont {Serga}, \citenamefont {Schneider}, \citenamefont
  {Leven},\ and\ \citenamefont {Hillebrands}}]{Kostylev2005}%
  \BibitemOpen
  \bibfield  {author} {\bibinfo {author} {\bibfnamefont {M.~P.}\ \bibnamefont
  {Kostylev}}, \bibinfo {author} {\bibfnamefont {A.~A.}\ \bibnamefont {Serga}},
  \bibinfo {author} {\bibfnamefont {T.}~\bibnamefont {Schneider}}, \bibinfo
  {author} {\bibfnamefont {B.}~\bibnamefont {Leven}}, \ and\ \bibinfo {author}
  {\bibfnamefont {B.}~\bibnamefont {Hillebrands}},\ }\href@noop {} {\bibfield
  {journal} {\bibinfo  {journal} {Applied Physics Letters}\ }\textbf {\bibinfo
  {volume} {87}},\ \bibinfo {eid} {153501} (\bibinfo {year}
  {2005})}\BibitemShut {NoStop}%
\bibitem [{\citenamefont {Khitun}\ and\ \citenamefont
  {Wang}(2005)}]{Khitun2005}%
  \BibitemOpen
  \bibfield  {author} {\bibinfo {author} {\bibfnamefont {A.}~\bibnamefont
  {Khitun}}\ and\ \bibinfo {author} {\bibfnamefont {K.~L.}\ \bibnamefont
  {Wang}},\ }\href@noop {} {\bibfield  {journal} {\bibinfo  {journal}
  {Superlattices and Microstructures}\ }\textbf {\bibinfo {volume} {38}},\
  \bibinfo {pages} {184 } (\bibinfo {year} {2005})}\BibitemShut {NoStop}%
\bibitem [{\citenamefont {Chumak}, \citenamefont {Serga},\ and\ \citenamefont
  {Hillebrands}(2013)}]{Chumak2013}%
  \BibitemOpen
  \bibfield  {author} {\bibinfo {author} {\bibfnamefont {A.~V.}\ \bibnamefont
  {Chumak}}, \bibinfo {author} {\bibfnamefont {A.~A.}\ \bibnamefont {Serga}}, \
  and\ \bibinfo {author} {\bibfnamefont {B.}~\bibnamefont {Hillebrands}},\
  }\href@noop {} {\bibfield  {journal} {\bibinfo  {journal} {Nature
  Communications}\ }\textbf {\bibinfo {volume} {5}} (\bibinfo {year}
  {2013})}\BibitemShut {NoStop}%
\bibitem [{\citenamefont {Demokritov}\ and\ \citenamefont
  {Slavin}(2013)}]{Demokritov2013}%
  \BibitemOpen
  \bibinfo {editor} {\bibfnamefont {S.}~\bibnamefont {Demokritov}}\ and\
  \bibinfo {editor} {\bibfnamefont {A.}~\bibnamefont {Slavin}},\ eds.,\
  \href@noop {} {\emph {\bibinfo {title} {Magnonics From Fundamentals to
  Applications}}}\ (\bibinfo  {publisher} {Springer},\ \bibinfo {year}
  {2013})\BibitemShut {NoStop}%
\bibitem [{\citenamefont {Krawczyk}\ and\ \citenamefont
  {Grundler}(2014)}]{Krawczyk2014}%
  \BibitemOpen
  \bibfield  {author} {\bibinfo {author} {\bibfnamefont {M.}~\bibnamefont
  {Krawczyk}}\ and\ \bibinfo {author} {\bibfnamefont {D.}~\bibnamefont
  {Grundler}},\ }\href@noop {} {\bibfield  {journal} {\bibinfo  {journal}
  {Journal of Physics: Condensed Matter}\ }\textbf {\bibinfo {volume} {26}},\
  \bibinfo {pages} {123202} (\bibinfo {year} {2014})}\BibitemShut {NoStop}%
\bibitem [{\citenamefont {Nikitov}, \citenamefont {Tailhades},\ and\
  \citenamefont {Tsai}(2001)}]{Nikitov2001}%
  \BibitemOpen
  \bibfield  {author} {\bibinfo {author} {\bibfnamefont {S.}~\bibnamefont
  {Nikitov}}, \bibinfo {author} {\bibfnamefont {P.}~\bibnamefont {Tailhades}},
  \ and\ \bibinfo {author} {\bibfnamefont {C.}~\bibnamefont {Tsai}},\
  }\href@noop {} {\bibfield  {journal} {\bibinfo  {journal} {Journal of
  Magnetism and Magnetic Materials}\ }\textbf {\bibinfo {volume} {236}},\
  \bibinfo {pages} {320 } (\bibinfo {year} {2001})}\BibitemShut {NoStop}%
\bibitem [{\citenamefont {Neusser}\ and\ \citenamefont
  {Grundler}(2009)}]{Neusser2009}%
  \BibitemOpen
  \bibfield  {author} {\bibinfo {author} {\bibfnamefont {S.}~\bibnamefont
  {Neusser}}\ and\ \bibinfo {author} {\bibfnamefont {D.}~\bibnamefont
  {Grundler}},\ }\href@noop {} {\bibfield  {journal} {\bibinfo  {journal}
  {Advanced Materials}\ }\textbf {\bibinfo {volume} {21}},\ \bibinfo {pages}
  {2927} (\bibinfo {year} {2009})}\BibitemShut {NoStop}%
\bibitem [{\citenamefont {Wang}\ \emph {et~al.}(2010)\citenamefont {Wang},
  \citenamefont {Zhang}, \citenamefont {Lim}, \citenamefont {Ng}, \citenamefont
  {Kuok}, \citenamefont {Jain},\ and\ \citenamefont {Adeyeye}}]{Wang2010}%
  \BibitemOpen
  \bibfield  {author} {\bibinfo {author} {\bibfnamefont {Z.}~\bibnamefont
  {Wang}}, \bibinfo {author} {\bibfnamefont {V.}~\bibnamefont {Zhang}},
  \bibinfo {author} {\bibfnamefont {H.}~\bibnamefont {Lim}}, \bibinfo {author}
  {\bibfnamefont {S.}~\bibnamefont {Ng}}, \bibinfo {author} {\bibfnamefont
  {M.}~\bibnamefont {Kuok}}, \bibinfo {author} {\bibfnamefont {S.}~\bibnamefont
  {Jain}}, \ and\ \bibinfo {author} {\bibfnamefont {A.}~\bibnamefont
  {Adeyeye}},\ }\href@noop {} {\bibfield  {journal} {\bibinfo  {journal} {ACS
  Nano}\ }\textbf {\bibinfo {volume} {4}},\ \bibinfo {pages} {643} (\bibinfo
  {year} {2010})}\BibitemShut {NoStop}%
\bibitem [{\citenamefont {Tacchi}\ \emph {et~al.}(2011)\citenamefont {Tacchi},
  \citenamefont {Montoncello}, \citenamefont {Madami}, \citenamefont
  {Gubbiotti}, \citenamefont {Carlotti}, \citenamefont {Giovannini},
  \citenamefont {Zivieri}, \citenamefont {Nizzoli}, \citenamefont {Jain},
  \citenamefont {Adeyeye},\ and\ \citenamefont {Singh}}]{Tacchi2011}%
  \BibitemOpen
  \bibfield  {author} {\bibinfo {author} {\bibfnamefont {S.}~\bibnamefont
  {Tacchi}}, \bibinfo {author} {\bibfnamefont {F.}~\bibnamefont {Montoncello}},
  \bibinfo {author} {\bibfnamefont {M.}~\bibnamefont {Madami}}, \bibinfo
  {author} {\bibfnamefont {G.}~\bibnamefont {Gubbiotti}}, \bibinfo {author}
  {\bibfnamefont {G.}~\bibnamefont {Carlotti}}, \bibinfo {author}
  {\bibfnamefont {L.}~\bibnamefont {Giovannini}}, \bibinfo {author}
  {\bibfnamefont {R.}~\bibnamefont {Zivieri}}, \bibinfo {author} {\bibfnamefont
  {F.}~\bibnamefont {Nizzoli}}, \bibinfo {author} {\bibfnamefont
  {S.}~\bibnamefont {Jain}}, \bibinfo {author} {\bibfnamefont {A.~O.}\
  \bibnamefont {Adeyeye}}, \ and\ \bibinfo {author} {\bibfnamefont
  {N.}~\bibnamefont {Singh}},\ }\href {\doibase 10.1103/PhysRevLett.107.127204}
  {\bibfield  {journal} {\bibinfo  {journal} {Phys. Rev. Lett.}\ }\textbf
  {\bibinfo {volume} {107}},\ \bibinfo {pages} {127204} (\bibinfo {year}
  {2011})}\BibitemShut {NoStop}%
\bibitem [{\citenamefont {Tacchi}\ \emph {et~al.}(2012)\citenamefont {Tacchi},
  \citenamefont {Duerr}, \citenamefont {Klos}, \citenamefont {Madami},
  \citenamefont {Neusser}, \citenamefont {Gubbiotti}, \citenamefont {Carlotti},
  \citenamefont {Krawczyk},\ and\ \citenamefont {Grundler}}]{Tacchi2012}%
  \BibitemOpen
  \bibfield  {author} {\bibinfo {author} {\bibfnamefont {S.}~\bibnamefont
  {Tacchi}}, \bibinfo {author} {\bibfnamefont {G.}~\bibnamefont {Duerr}},
  \bibinfo {author} {\bibfnamefont {J.~W.}\ \bibnamefont {Klos}}, \bibinfo
  {author} {\bibfnamefont {M.}~\bibnamefont {Madami}}, \bibinfo {author}
  {\bibfnamefont {S.}~\bibnamefont {Neusser}}, \bibinfo {author} {\bibfnamefont
  {G.}~\bibnamefont {Gubbiotti}}, \bibinfo {author} {\bibfnamefont
  {G.}~\bibnamefont {Carlotti}}, \bibinfo {author} {\bibfnamefont
  {M.}~\bibnamefont {Krawczyk}}, \ and\ \bibinfo {author} {\bibfnamefont
  {D.}~\bibnamefont {Grundler}},\ }\href@noop {} {\bibfield  {journal}
  {\bibinfo  {journal} {Phys. Rev. Lett.}\ }\textbf {\bibinfo {volume} {109}},\
  \bibinfo {pages} {137202} (\bibinfo {year} {2012})}\BibitemShut {NoStop}%
\bibitem [{\citenamefont {Kruglyak}, \citenamefont {Demokritov},\ and\
  \citenamefont {Grundler}(2010)}]{Kruglyak2010}%
  \BibitemOpen
  \bibfield  {author} {\bibinfo {author} {\bibfnamefont {V.~V.}\ \bibnamefont
  {Kruglyak}}, \bibinfo {author} {\bibfnamefont {S.~O.}\ \bibnamefont
  {Demokritov}}, \ and\ \bibinfo {author} {\bibfnamefont {D.}~\bibnamefont
  {Grundler}},\ }\href@noop {} {\bibfield  {journal} {\bibinfo  {journal} {J.
  Phys. D: Appl. Phys.}\ }\textbf {\bibinfo {volume} {43}},\ \bibinfo {pages}
  {264001} (\bibinfo {year} {2010})}\BibitemShut {NoStop}%
\bibitem [{\citenamefont {Lenk}\ \emph {et~al.}(2011)\citenamefont {Lenk},
  \citenamefont {Ulrichs}, \citenamefont {Garbs},\ and\ \citenamefont
  {M\"unzenberg}}]{Lenk2011}%
  \BibitemOpen
  \bibfield  {author} {\bibinfo {author} {\bibfnamefont {B.}~\bibnamefont
  {Lenk}}, \bibinfo {author} {\bibfnamefont {H.}~\bibnamefont {Ulrichs}},
  \bibinfo {author} {\bibfnamefont {F.}~\bibnamefont {Garbs}}, \ and\ \bibinfo
  {author} {\bibfnamefont {M.}~\bibnamefont {M\"unzenberg}},\ }\href@noop {}
  {\bibfield  {journal} {\bibinfo  {journal} {Physics Reports}\ }\textbf
  {\bibinfo {volume} {507}},\ \bibinfo {pages} {107 } (\bibinfo {year}
  {2011})}\BibitemShut {NoStop}%
\bibitem [{\citenamefont {Sklenar}\ \emph {et~al.}(2015)\citenamefont
  {Sklenar}, \citenamefont {Tucciarone}, \citenamefont {Lee}, \citenamefont
  {Tice}, \citenamefont {Chang}, \citenamefont {Lee}, \citenamefont
  {Nevirkovets}, \citenamefont {Heinonen},\ and\ \citenamefont
  {Ketterson}}]{Sklenar2015}%
  \BibitemOpen
  \bibfield  {author} {\bibinfo {author} {\bibfnamefont {J.}~\bibnamefont
  {Sklenar}}, \bibinfo {author} {\bibfnamefont {P.}~\bibnamefont {Tucciarone}},
  \bibinfo {author} {\bibfnamefont {R.~J.}\ \bibnamefont {Lee}}, \bibinfo
  {author} {\bibfnamefont {D.}~\bibnamefont {Tice}}, \bibinfo {author}
  {\bibfnamefont {R.~P.~H.}\ \bibnamefont {Chang}}, \bibinfo {author}
  {\bibfnamefont {S.~J.}\ \bibnamefont {Lee}}, \bibinfo {author} {\bibfnamefont
  {I.~P.}\ \bibnamefont {Nevirkovets}}, \bibinfo {author} {\bibfnamefont
  {O.}~\bibnamefont {Heinonen}}, \ and\ \bibinfo {author} {\bibfnamefont
  {J.~B.}\ \bibnamefont {Ketterson}},\ }\href@noop {} {\bibfield  {journal}
  {\bibinfo  {journal} {Phys. Rev. B}\ }\textbf {\bibinfo {volume} {91}},\
  \bibinfo {pages} {134424} (\bibinfo {year} {2015})}\BibitemShut {NoStop}%
\bibitem [{\citenamefont {Grundler}(2015)}]{Grundler2015}%
  \BibitemOpen
  \bibfield  {author} {\bibinfo {author} {\bibfnamefont {D.}~\bibnamefont
  {Grundler}},\ }\href@noop {} {\bibfield  {journal} {\bibinfo  {journal}
  {Nature Physics}\ }\textbf {\bibinfo {volume} {11}},\ \bibinfo {pages} {438 }
  (\bibinfo {year} {2015})}\BibitemShut {NoStop}%
\bibitem [{\citenamefont {Karenowska}\ \emph {et~al.}(2012)\citenamefont
  {Karenowska}, \citenamefont {Gregg}, \citenamefont {Tiberkevich},
  \citenamefont {Slavin}, \citenamefont {Chumak}, \citenamefont {Serga},\ and\
  \citenamefont {Hillebrands}}]{Karenowska2012}%
  \BibitemOpen
  \bibfield  {author} {\bibinfo {author} {\bibfnamefont {A.~D.}\ \bibnamefont
  {Karenowska}}, \bibinfo {author} {\bibfnamefont {J.~F.}\ \bibnamefont
  {Gregg}}, \bibinfo {author} {\bibfnamefont {V.~S.}\ \bibnamefont
  {Tiberkevich}}, \bibinfo {author} {\bibfnamefont {A.~N.}\ \bibnamefont
  {Slavin}}, \bibinfo {author} {\bibfnamefont {A.~V.}\ \bibnamefont {Chumak}},
  \bibinfo {author} {\bibfnamefont {A.~A.}\ \bibnamefont {Serga}}, \ and\
  \bibinfo {author} {\bibfnamefont {B.}~\bibnamefont {Hillebrands}},\ }\href
  {\doibase 10.1103/PhysRevLett.108.015505} {\bibfield  {journal} {\bibinfo
  {journal} {Phys. Rev. Lett.}\ }\textbf {\bibinfo {volume} {108}},\ \bibinfo
  {pages} {015505} (\bibinfo {year} {2012})}\BibitemShut {NoStop}%
\bibitem [{\citenamefont {Vogel}\ \emph {et~al.}(2015)\citenamefont {Vogel},
  \citenamefont {Chumak}, \citenamefont {Waller}, \citenamefont {Langner},
  \citenamefont {Vasyuchka}, \citenamefont {Hillebrands},\ and\ \citenamefont
  {von Freymann}}]{Vogel2010}%
  \BibitemOpen
  \bibfield  {author} {\bibinfo {author} {\bibfnamefont {M.}~\bibnamefont
  {Vogel}}, \bibinfo {author} {\bibfnamefont {A.~V.}\ \bibnamefont {Chumak}},
  \bibinfo {author} {\bibfnamefont {E.~H.}\ \bibnamefont {Waller}}, \bibinfo
  {author} {\bibfnamefont {T.}~\bibnamefont {Langner}}, \bibinfo {author}
  {\bibfnamefont {V.~I.}\ \bibnamefont {Vasyuchka}}, \bibinfo {author}
  {\bibfnamefont {B.}~\bibnamefont {Hillebrands}}, \ and\ \bibinfo {author}
  {\bibfnamefont {G.}~\bibnamefont {von Freymann}},\ }\href@noop {} {\bibfield
  {journal} {\bibinfo  {journal} {Nature Physics}\ }\textbf {\bibinfo {volume}
  {11}} (\bibinfo {year} {2015})}\BibitemShut {NoStop}%
\bibitem [{\citenamefont {Wang}\ \emph {et~al.}(2005)\citenamefont {Wang},
  \citenamefont {Nisoli}, \citenamefont {Freitas}, \citenamefont {Li},
  \citenamefont {McConville}, \citenamefont {Cooley}, \citenamefont {Lund},
  \citenamefont {Samarth}, \citenamefont {Leighton}, \citenamefont {Crespi},\
  and\ \citenamefont {Schiffer}}]{Wang2005}%
  \BibitemOpen
  \bibfield  {author} {\bibinfo {author} {\bibfnamefont {R.}~\bibnamefont
  {Wang}}, \bibinfo {author} {\bibfnamefont {C.}~\bibnamefont {Nisoli}},
  \bibinfo {author} {\bibfnamefont {R.}~\bibnamefont {Freitas}}, \bibinfo
  {author} {\bibfnamefont {J.}~\bibnamefont {Li}}, \bibinfo {author}
  {\bibfnamefont {W.}~\bibnamefont {McConville}}, \bibinfo {author}
  {\bibfnamefont {B.}~\bibnamefont {Cooley}}, \bibinfo {author} {\bibfnamefont
  {M.}~\bibnamefont {Lund}}, \bibinfo {author} {\bibfnamefont {N.}~\bibnamefont
  {Samarth}}, \bibinfo {author} {\bibfnamefont {C.}~\bibnamefont {Leighton}},
  \bibinfo {author} {\bibfnamefont {V.}~\bibnamefont {Crespi}}, \ and\ \bibinfo
  {author} {\bibfnamefont {P.}~\bibnamefont {Schiffer}},\ }\href@noop {}
  {\bibfield  {journal} {\bibinfo  {journal} {Nature}\ }\textbf {\bibinfo
  {volume} {439}},\ \bibinfo {pages} {303} (\bibinfo {year}
  {2005})}\BibitemShut {NoStop}%
\bibitem [{\citenamefont {Nisoli}, \citenamefont {Moessner},\ and\
  \citenamefont {Schiffer}(2013)}]{Nisoli2013}%
  \BibitemOpen
  \bibfield  {author} {\bibinfo {author} {\bibfnamefont {C.}~\bibnamefont
  {Nisoli}}, \bibinfo {author} {\bibfnamefont {R.}~\bibnamefont {Moessner}}, \
  and\ \bibinfo {author} {\bibfnamefont {P.}~\bibnamefont {Schiffer}},\
  }\href@noop {} {\bibfield  {journal} {\bibinfo  {journal} {Rev. Mod. Phys.}\
  }\textbf {\bibinfo {volume} {85}},\ \bibinfo {pages} {1473} (\bibinfo {year}
  {2013})}\BibitemShut {NoStop}%
\bibitem [{\citenamefont {Heyderman}\ and\ \citenamefont
  {Stamps}(2013)}]{Heyderman2013}%
  \BibitemOpen
  \bibfield  {author} {\bibinfo {author} {\bibfnamefont {L.~J.}\ \bibnamefont
  {Heyderman}}\ and\ \bibinfo {author} {\bibfnamefont {R.~L.}\ \bibnamefont
  {Stamps}},\ }\href@noop {} {\bibfield  {journal} {\bibinfo  {journal}
  {Journal of Physics: Condensed Matter}\ }\textbf {\bibinfo {volume} {25}},\
  \bibinfo {pages} {363201} (\bibinfo {year} {2013})}\BibitemShut {NoStop}%
\bibitem [{\citenamefont {Stamps}\ \emph {et~al.}(2014)\citenamefont {Stamps},
  \citenamefont {Breitkreutz}, \citenamefont {\AA{}kerman}, \citenamefont
  {Chumak}, \citenamefont {Otani}, \citenamefont {Bauer}, \citenamefont
  {Thiele}, \citenamefont {Bowen}, \citenamefont {Majetich}, \citenamefont
  {Kl\"{a}ui}, \citenamefont {Prejbeanu}, \citenamefont {Dieny}, \citenamefont
  {Dempsey},\ and\ \citenamefont {Hillebrands}}]{Roadmap2014}%
  \BibitemOpen
  \bibfield  {author} {\bibinfo {author} {\bibfnamefont {R.~L.}\ \bibnamefont
  {Stamps}}, \bibinfo {author} {\bibfnamefont {S.}~\bibnamefont {Breitkreutz}},
  \bibinfo {author} {\bibfnamefont {J.}~\bibnamefont {\AA{}kerman}}, \bibinfo
  {author} {\bibfnamefont {A.~V.}\ \bibnamefont {Chumak}}, \bibinfo {author}
  {\bibfnamefont {Y.}~\bibnamefont {Otani}}, \bibinfo {author} {\bibfnamefont
  {G.~E.~W.}\ \bibnamefont {Bauer}}, \bibinfo {author} {\bibfnamefont {J.-U.}\
  \bibnamefont {Thiele}}, \bibinfo {author} {\bibfnamefont {M.}~\bibnamefont
  {Bowen}}, \bibinfo {author} {\bibfnamefont {S.~A.}\ \bibnamefont {Majetich}},
  \bibinfo {author} {\bibfnamefont {M.}~\bibnamefont {Kl\"{a}ui}}, \bibinfo
  {author} {\bibfnamefont {I.~L.}\ \bibnamefont {Prejbeanu}}, \bibinfo {author}
  {\bibfnamefont {B.}~\bibnamefont {Dieny}}, \bibinfo {author} {\bibfnamefont
  {N.~M.}\ \bibnamefont {Dempsey}}, \ and\ \bibinfo {author} {\bibfnamefont
  {B.}~\bibnamefont {Hillebrands}},\ }\href@noop {} {\bibfield  {journal}
  {\bibinfo  {journal} {Journal of Physics D: Applied Physics}\ }\textbf
  {\bibinfo {volume} {47}},\ \bibinfo {pages} {33} (\bibinfo {year}
  {2014})}\BibitemShut {NoStop}%
\bibitem [{\citenamefont {Kapaklis}\ \emph {et~al.}(2014)\citenamefont
  {Kapaklis}, \citenamefont {Arnalds}, \citenamefont {Farhan}, \citenamefont
  {Chopdekar}, \citenamefont {Balan}, \citenamefont {Scholl}, \citenamefont
  {Heyderman},\ and\ \citenamefont {Hj\"{o}rvarsson}}]{Kapaklis2014}%
  \BibitemOpen
  \bibfield  {author} {\bibinfo {author} {\bibfnamefont {V.}~\bibnamefont
  {Kapaklis}}, \bibinfo {author} {\bibfnamefont {U.}~\bibnamefont {Arnalds}},
  \bibinfo {author} {\bibfnamefont {A.}~\bibnamefont {Farhan}}, \bibinfo
  {author} {\bibfnamefont {R.}~\bibnamefont {Chopdekar}}, \bibinfo {author}
  {\bibfnamefont {A.}~\bibnamefont {Balan}}, \bibinfo {author} {\bibfnamefont
  {A.}~\bibnamefont {Scholl}}, \bibinfo {author} {\bibfnamefont
  {L.}~\bibnamefont {Heyderman}}, \ and\ \bibinfo {author} {\bibfnamefont
  {B.}~\bibnamefont {Hj\"{o}rvarsson}},\ }\href@noop {} {\bibfield  {journal}
  {\bibinfo  {journal} {Nature Nanotechnology}\ } (\bibinfo {year}
  {2014})}\BibitemShut {NoStop}%
\bibitem [{\citenamefont {Qi}, \citenamefont {Brintlinger},\ and\ \citenamefont
  {Cumings}(2008)}]{Qi2008}%
  \BibitemOpen
  \bibfield  {author} {\bibinfo {author} {\bibfnamefont {Y.}~\bibnamefont
  {Qi}}, \bibinfo {author} {\bibfnamefont {T.}~\bibnamefont {Brintlinger}}, \
  and\ \bibinfo {author} {\bibfnamefont {J.}~\bibnamefont {Cumings}},\
  }\href@noop {} {\bibfield  {journal} {\bibinfo  {journal} {Phys. Rev. B}\
  }\textbf {\bibinfo {volume} {77}},\ \bibinfo {pages} {094418} (\bibinfo
  {year} {2008})}\BibitemShut {NoStop}%
\bibitem [{\citenamefont {Gliga}\ \emph {et~al.}(2013)\citenamefont {Gliga},
  \citenamefont {K\'akay}, \citenamefont {Hertel},\ and\ \citenamefont
  {Heinonen}}]{Gliga2013}%
  \BibitemOpen
  \bibfield  {author} {\bibinfo {author} {\bibfnamefont {S.}~\bibnamefont
  {Gliga}}, \bibinfo {author} {\bibfnamefont {A.}~\bibnamefont {K\'akay}},
  \bibinfo {author} {\bibfnamefont {R.}~\bibnamefont {Hertel}}, \ and\ \bibinfo
  {author} {\bibfnamefont {O.~G.}\ \bibnamefont {Heinonen}},\ }\href@noop {}
  {\bibfield  {journal} {\bibinfo  {journal} {Phys. Rev. Lett.}\ }\textbf
  {\bibinfo {volume} {110}},\ \bibinfo {pages} {117205} (\bibinfo {year}
  {2013})}\BibitemShut {NoStop}%
\bibitem [{\citenamefont {Gliga}\ \emph {et~al.}(2015)\citenamefont {Gliga},
  \citenamefont {K\'akay}, \citenamefont {Heyderman}, \citenamefont {Hertel},\
  and\ \citenamefont {Heinonen}}]{Gliga2015}%
  \BibitemOpen
  \bibfield  {author} {\bibinfo {author} {\bibfnamefont {S.}~\bibnamefont
  {Gliga}}, \bibinfo {author} {\bibfnamefont {A.}~\bibnamefont {K\'akay}},
  \bibinfo {author} {\bibfnamefont {L.~J.}\ \bibnamefont {Heyderman}}, \bibinfo
  {author} {\bibfnamefont {R.}~\bibnamefont {Hertel}}, \ and\ \bibinfo {author}
  {\bibfnamefont {O.~G.}\ \bibnamefont {Heinonen}},\ }\href@noop {} {\bibfield
  {journal} {\bibinfo  {journal} {Phys. Rev. B}\ }\textbf {\bibinfo {volume}
  {92}},\ \bibinfo {pages} {060413} (\bibinfo {year} {2015})}\BibitemShut
  {NoStop}%
\bibitem [{\citenamefont {Balkanski}\ and\ \citenamefont
  {Wallis}(2000)}]{Balkanski2000}%
  \BibitemOpen
  \bibfield  {author} {\bibinfo {author} {\bibfnamefont {M.}~\bibnamefont
  {Balkanski}}\ and\ \bibinfo {author} {\bibfnamefont {R.}~\bibnamefont
  {Wallis}},\ }\href@noop {} {\emph {\bibinfo {title} {Semiconductor physics
  and applications}}}\ (\bibinfo  {publisher} {Oxford University Press},\
  \bibinfo {year} {2000})\BibitemShut {NoStop}%
\bibitem [{\citenamefont {Jungfleisch}\ \emph {et~al.}(2016)\citenamefont
  {Jungfleisch}, \citenamefont {Zhang}, \citenamefont {Iacocca}, \citenamefont
  {Sklenar}, \citenamefont {Ding}, \citenamefont {Jiang}, \citenamefont
  {Zhang}, \citenamefont {Pearson}, \citenamefont {Novosad}, \citenamefont
  {Ketterson}, \citenamefont {Heinonen},\ and\ \citenamefont
  {Hoffmann}}]{Jungfleisch2016}%
  \BibitemOpen
  \bibfield  {author} {\bibinfo {author} {\bibfnamefont {M.~B.}\ \bibnamefont
  {Jungfleisch}}, \bibinfo {author} {\bibfnamefont {W.}~\bibnamefont {Zhang}},
  \bibinfo {author} {\bibfnamefont {E.}~\bibnamefont {Iacocca}}, \bibinfo
  {author} {\bibfnamefont {J.}~\bibnamefont {Sklenar}}, \bibinfo {author}
  {\bibfnamefont {J.}~\bibnamefont {Ding}}, \bibinfo {author} {\bibfnamefont
  {W.}~\bibnamefont {Jiang}}, \bibinfo {author} {\bibfnamefont
  {S.}~\bibnamefont {Zhang}}, \bibinfo {author} {\bibfnamefont {J.~E.}\
  \bibnamefont {Pearson}}, \bibinfo {author} {\bibfnamefont {V.}~\bibnamefont
  {Novosad}}, \bibinfo {author} {\bibfnamefont {J.~B.}\ \bibnamefont
  {Ketterson}}, \bibinfo {author} {\bibfnamefont {O.}~\bibnamefont {Heinonen}},
  \ and\ \bibinfo {author} {\bibfnamefont {A.}~\bibnamefont {Hoffmann}},\
  }\href {\doibase 10.1103/PhysRevB.93.100401} {\bibfield  {journal} {\bibinfo
  {journal} {Phys. Rev. B}\ }\textbf {\bibinfo {volume} {93}},\ \bibinfo
  {pages} {100401} (\bibinfo {year} {2016})}\BibitemShut {NoStop}%
\bibitem [{\citenamefont {Bhat}\ \emph {et~al.}(2016)\citenamefont {Bhat},
  \citenamefont {Heimbach}, \citenamefont {Stasinopoulos},\ and\ \citenamefont
  {Grundler}}]{Bhat2016}%
  \BibitemOpen
  \bibfield  {author} {\bibinfo {author} {\bibfnamefont {V.~S.}\ \bibnamefont
  {Bhat}}, \bibinfo {author} {\bibfnamefont {F.}~\bibnamefont {Heimbach}},
  \bibinfo {author} {\bibfnamefont {I.}~\bibnamefont {Stasinopoulos}}, \ and\
  \bibinfo {author} {\bibfnamefont {D.}~\bibnamefont {Grundler}},\ }\href@noop
  {} {\ \textbf {\bibinfo {volume} {arXiv:1602.00918}} (\bibinfo {year}
  {2016})}\BibitemShut {NoStop}%
\bibitem [{\citenamefont {Zhou}\ \emph {et~al.}(2016)\citenamefont {Zhou},
  \citenamefont {Chua}, \citenamefont {Singh},\ and\ \citenamefont
  {Adeyeye}}]{Zhou2016}%
  \BibitemOpen
  \bibfield  {author} {\bibinfo {author} {\bibfnamefont {X.}~\bibnamefont
  {Zhou}}, \bibinfo {author} {\bibfnamefont {G.-L.}\ \bibnamefont {Chua}},
  \bibinfo {author} {\bibfnamefont {N.}~\bibnamefont {Singh}}, \ and\ \bibinfo
  {author} {\bibfnamefont {A.~O.}\ \bibnamefont {Adeyeye}},\ }\href {\doibase
  10.1002/adfm.201505165} {\bibfield  {journal} {\bibinfo  {journal} {Adv.
  Func. Mater.}\ } (\bibinfo {year} {2016}),\
  10.1002/adfm.201505165}\BibitemShut {NoStop}%
\bibitem [{\citenamefont {Shindou}\ \emph
  {et~al.}(2013{\natexlab{a}})\citenamefont {Shindou}, \citenamefont
  {Matsumoto}, \citenamefont {Murakami},\ and\ \citenamefont
  {Ohe}}]{Shindou2013}%
  \BibitemOpen
  \bibfield  {author} {\bibinfo {author} {\bibfnamefont {R.}~\bibnamefont
  {Shindou}}, \bibinfo {author} {\bibfnamefont {R.}~\bibnamefont {Matsumoto}},
  \bibinfo {author} {\bibfnamefont {S.}~\bibnamefont {Murakami}}, \ and\
  \bibinfo {author} {\bibfnamefont {J.-i.}\ \bibnamefont {Ohe}},\ }\href@noop
  {} {\bibfield  {journal} {\bibinfo  {journal} {Phys. Rev. B}\ }\textbf
  {\bibinfo {volume} {87}},\ \bibinfo {pages} {174427} (\bibinfo {year}
  {2013}{\natexlab{a}})}\BibitemShut {NoStop}%
\bibitem [{\citenamefont {Shindou}\ \emph
  {et~al.}(2013{\natexlab{b}})\citenamefont {Shindou}, \citenamefont {Ohe},
  \citenamefont {Matsumoto}, \citenamefont {Murakami},\ and\ \citenamefont
  {Saitoh}}]{Shindou2013b}%
  \BibitemOpen
  \bibfield  {author} {\bibinfo {author} {\bibfnamefont {R.}~\bibnamefont
  {Shindou}}, \bibinfo {author} {\bibfnamefont {J.-i.}\ \bibnamefont {Ohe}},
  \bibinfo {author} {\bibfnamefont {R.}~\bibnamefont {Matsumoto}}, \bibinfo
  {author} {\bibfnamefont {S.}~\bibnamefont {Murakami}}, \ and\ \bibinfo
  {author} {\bibfnamefont {E.}~\bibnamefont {Saitoh}},\ }\href@noop {}
  {\bibfield  {journal} {\bibinfo  {journal} {Phys. Rev. B}\ }\textbf {\bibinfo
  {volume} {87}},\ \bibinfo {pages} {174402} (\bibinfo {year}
  {2013}{\natexlab{b}})}\BibitemShut {NoStop}%
\bibitem [{\citenamefont {Farhan}\ \emph {et~al.}(2013)\citenamefont {Farhan},
  \citenamefont {Derlet}, \citenamefont {Kleibert}, \citenamefont {Balan},
  \citenamefont {Chopdekar}, \citenamefont {Wyss}, \citenamefont {Perron},
  \citenamefont {Scholl}, \citenamefont {Nolting},\ and\ \citenamefont
  {Heyderman}}]{Farhan2013}%
  \BibitemOpen
  \bibfield  {author} {\bibinfo {author} {\bibfnamefont {A.}~\bibnamefont
  {Farhan}}, \bibinfo {author} {\bibfnamefont {P.~M.}\ \bibnamefont {Derlet}},
  \bibinfo {author} {\bibfnamefont {A.}~\bibnamefont {Kleibert}}, \bibinfo
  {author} {\bibfnamefont {A.}~\bibnamefont {Balan}}, \bibinfo {author}
  {\bibfnamefont {R.~V.}\ \bibnamefont {Chopdekar}}, \bibinfo {author}
  {\bibfnamefont {M.}~\bibnamefont {Wyss}}, \bibinfo {author} {\bibfnamefont
  {J.}~\bibnamefont {Perron}}, \bibinfo {author} {\bibfnamefont
  {A.}~\bibnamefont {Scholl}}, \bibinfo {author} {\bibfnamefont
  {F.}~\bibnamefont {Nolting}}, \ and\ \bibinfo {author} {\bibfnamefont
  {L.~J.}\ \bibnamefont {Heyderman}},\ }\href@noop {} {\bibfield  {journal}
  {\bibinfo  {journal} {Phys. Rev. Lett.}\ }\textbf {\bibinfo {volume} {111}},\
  \bibinfo {pages} {057204} (\bibinfo {year} {2013})}\BibitemShut {NoStop}%
\bibitem [{\citenamefont {Cowburn}(2000)}]{Cowburn2000}%
  \BibitemOpen
  \bibfield  {author} {\bibinfo {author} {\bibfnamefont {R.~P.}\ \bibnamefont
  {Cowburn}},\ }\href@noop {} {\bibfield  {journal} {\bibinfo  {journal}
  {Journal of Physics D: Applied Physics}\ }\textbf {\bibinfo {volume} {33}},\
  \bibinfo {pages} {R1} (\bibinfo {year} {2000})}\BibitemShut {NoStop}%
\bibitem [{\citenamefont {Madami}\ \emph {et~al.}(2011)\citenamefont {Madami},
  \citenamefont {Carlotti}, \citenamefont {Gubbiotti}, \citenamefont
  {Scarponi}, \citenamefont {Tacchi},\ and\ \citenamefont {Ono}}]{Madami2011b}%
  \BibitemOpen
  \bibfield  {author} {\bibinfo {author} {\bibfnamefont {M.}~\bibnamefont
  {Madami}}, \bibinfo {author} {\bibfnamefont {G.}~\bibnamefont {Carlotti}},
  \bibinfo {author} {\bibfnamefont {G.}~\bibnamefont {Gubbiotti}}, \bibinfo
  {author} {\bibfnamefont {F.}~\bibnamefont {Scarponi}}, \bibinfo {author}
  {\bibfnamefont {S.}~\bibnamefont {Tacchi}}, \ and\ \bibinfo {author}
  {\bibfnamefont {T.}~\bibnamefont {Ono}},\ }\href@noop {} {\bibfield
  {journal} {\bibinfo  {journal} {Journal of Applied Physics}\ }\textbf
  {\bibinfo {volume} {109}},\ \bibinfo {eid} {07B901} (\bibinfo {year}
  {2011})}\BibitemShut {NoStop}%
\bibitem [{\citenamefont {Carlotti}\ \emph {et~al.}(2014)\citenamefont
  {Carlotti}, \citenamefont {Gubbiotti}, \citenamefont {Madami}, \citenamefont
  {Tacchi}, \citenamefont {Hartmann}, \citenamefont {Emmerling}, \citenamefont
  {Kamp},\ and\ \citenamefont {Worschech}}]{Carlotti2014}%
  \BibitemOpen
  \bibfield  {author} {\bibinfo {author} {\bibfnamefont {G.}~\bibnamefont
  {Carlotti}}, \bibinfo {author} {\bibfnamefont {G.}~\bibnamefont {Gubbiotti}},
  \bibinfo {author} {\bibfnamefont {M.}~\bibnamefont {Madami}}, \bibinfo
  {author} {\bibfnamefont {S.}~\bibnamefont {Tacchi}}, \bibinfo {author}
  {\bibfnamefont {F.}~\bibnamefont {Hartmann}}, \bibinfo {author}
  {\bibfnamefont {M.}~\bibnamefont {Emmerling}}, \bibinfo {author}
  {\bibfnamefont {M.}~\bibnamefont {Kamp}}, \ and\ \bibinfo {author}
  {\bibfnamefont {L.}~\bibnamefont {Worschech}},\ }\href@noop {} {\bibfield
  {journal} {\bibinfo  {journal} {Journal of Physics D: Applied Physics}\
  }\textbf {\bibinfo {volume} {47}},\ \bibinfo {pages} {265001} (\bibinfo
  {year} {2014})}\BibitemShut {NoStop}%
\bibitem [{\citenamefont {Slavin}\ and\ \citenamefont
  {Tiberkevich}(2009)}]{Slavin2009}%
  \BibitemOpen
  \bibfield  {author} {\bibinfo {author} {\bibfnamefont {A.}~\bibnamefont
  {Slavin}}\ and\ \bibinfo {author} {\bibfnamefont {V.}~\bibnamefont
  {Tiberkevich}},\ }\href@noop {} {\bibfield  {journal} {\bibinfo  {journal}
  {Magnetics, IEEE Transactions on}\ }\textbf {\bibinfo {volume} {45}},\
  \bibinfo {pages} {1875 } (\bibinfo {year} {2009})}\BibitemShut {NoStop}%
\bibitem [{\citenamefont {Slavin}\ and\ \citenamefont
  {Tiberkevich}(2005)}]{Slavin2005}%
  \BibitemOpen
  \bibfield  {author} {\bibinfo {author} {\bibfnamefont {A.}~\bibnamefont
  {Slavin}}\ and\ \bibinfo {author} {\bibfnamefont {V.}~\bibnamefont
  {Tiberkevich}},\ }\href@noop {} {\bibfield  {journal} {\bibinfo  {journal}
  {Phys. Rev. Lett.}\ }\textbf {\bibinfo {volume} {95}},\ \bibinfo {pages}
  {237201} (\bibinfo {year} {2005})}\BibitemShut {NoStop}%
\bibitem [{\citenamefont {Bonetti}\ \emph {et~al.}(2010)\citenamefont
  {Bonetti}, \citenamefont {Tiberkevich}, \citenamefont {Consolo},
  \citenamefont {Finocchio}, \citenamefont {Muduli}, \citenamefont {Mancoff},
  \citenamefont {Slavin},\ and\ \citenamefont {\AA{}kerman}}]{Bonetti2010}%
  \BibitemOpen
  \bibfield  {author} {\bibinfo {author} {\bibfnamefont {S.}~\bibnamefont
  {Bonetti}}, \bibinfo {author} {\bibfnamefont {V.}~\bibnamefont
  {Tiberkevich}}, \bibinfo {author} {\bibfnamefont {G.}~\bibnamefont
  {Consolo}}, \bibinfo {author} {\bibfnamefont {G.}~\bibnamefont {Finocchio}},
  \bibinfo {author} {\bibfnamefont {P.}~\bibnamefont {Muduli}}, \bibinfo
  {author} {\bibfnamefont {F.}~\bibnamefont {Mancoff}}, \bibinfo {author}
  {\bibfnamefont {A.}~\bibnamefont {Slavin}}, \ and\ \bibinfo {author}
  {\bibfnamefont {J.}~\bibnamefont {\AA{}kerman}},\ }\href@noop {} {\bibfield
  {journal} {\bibinfo  {journal} {Phys. Rev. Lett.}\ }\textbf {\bibinfo
  {volume} {105}},\ \bibinfo {pages} {217204} (\bibinfo {year}
  {2010})}\BibitemShut {NoStop}%
\bibitem [{\citenamefont {Iacocca}\ and\ \citenamefont
  {\AA{}kerman}(2012)}]{Iacocca2012}%
  \BibitemOpen
  \bibfield  {author} {\bibinfo {author} {\bibfnamefont {E.}~\bibnamefont
  {Iacocca}}\ and\ \bibinfo {author} {\bibfnamefont {J.}~\bibnamefont
  {\AA{}kerman}},\ }\href@noop {} {\bibfield  {journal} {\bibinfo  {journal}
  {Phys. Rev. B}\ }\textbf {\bibinfo {volume} {85}},\ \bibinfo {pages} {184420}
  (\bibinfo {year} {2012})}\BibitemShut {NoStop}%
\bibitem [{\citenamefont {Iacocca}\ and\ \citenamefont
  {\AA{}kerman}(2013)}]{Iacocca2013b}%
  \BibitemOpen
  \bibfield  {author} {\bibinfo {author} {\bibfnamefont {E.}~\bibnamefont
  {Iacocca}}\ and\ \bibinfo {author} {\bibfnamefont {J.}~\bibnamefont
  {\AA{}kerman}},\ }\href@noop {} {\bibfield  {journal} {\bibinfo  {journal}
  {Phys. Rev. B}\ }\textbf {\bibinfo {volume} {87}},\ \bibinfo {pages} {214428}
  (\bibinfo {year} {2013})}\BibitemShut {NoStop}%
\bibitem [{\citenamefont {Iacocca}\ \emph {et~al.}(2014)\citenamefont
  {Iacocca}, \citenamefont {Heinonen}, \citenamefont {Muduli},\ and\
  \citenamefont {\AA{}kerman}}]{Iacocca2014b}%
  \BibitemOpen
  \bibfield  {author} {\bibinfo {author} {\bibfnamefont {E.}~\bibnamefont
  {Iacocca}}, \bibinfo {author} {\bibfnamefont {O.}~\bibnamefont {Heinonen}},
  \bibinfo {author} {\bibfnamefont {P.~K.}\ \bibnamefont {Muduli}}, \ and\
  \bibinfo {author} {\bibfnamefont {J.}~\bibnamefont {\AA{}kerman}},\
  }\href@noop {} {\bibfield  {journal} {\bibinfo  {journal} {Phys. Rev. B}\
  }\textbf {\bibinfo {volume} {89}},\ \bibinfo {pages} {054402} (\bibinfo
  {year} {2014})}\BibitemShut {NoStop}%
\bibitem [{\citenamefont {Iacocca}\ \emph {et~al.}(2015)\citenamefont
  {Iacocca}, \citenamefont {D\"urrenfeld}, \citenamefont {Heinonen},
  \citenamefont {\AA{}kerman},\ and\ \citenamefont {Dumas}}]{Iacocca2015}%
  \BibitemOpen
  \bibfield  {author} {\bibinfo {author} {\bibfnamefont {E.}~\bibnamefont
  {Iacocca}}, \bibinfo {author} {\bibfnamefont {P.}~\bibnamefont
  {D\"urrenfeld}}, \bibinfo {author} {\bibfnamefont {O.}~\bibnamefont
  {Heinonen}}, \bibinfo {author} {\bibfnamefont {J.}~\bibnamefont
  {\AA{}kerman}}, \ and\ \bibinfo {author} {\bibfnamefont {R.~K.}\ \bibnamefont
  {Dumas}},\ }\href@noop {} {\bibfield  {journal} {\bibinfo  {journal} {Phys.
  Rev. B}\ }\textbf {\bibinfo {volume} {91}},\ \bibinfo {pages} {104405}
  (\bibinfo {year} {2015})}\BibitemShut {NoStop}%
\bibitem [{\citenamefont {Locatelli}\ \emph {et~al.}(2015)\citenamefont
  {Locatelli}, \citenamefont {Hamadeh}, \citenamefont {Abreu~Araujo},
  \citenamefont {Belanovsky}, \citenamefont {Skirdkov}, \citenamefont {Lebrun},
  \citenamefont {Naletov}, \citenamefont {Zvezdin}, \citenamefont {M},
  \citenamefont {Grollier}, \citenamefont {Klein}, \citenamefont {Cros},\ and\
  \citenamefont {de~Loubens}}]{Locatelli2015}%
  \BibitemOpen
  \bibfield  {author} {\bibinfo {author} {\bibfnamefont {N.}~\bibnamefont
  {Locatelli}}, \bibinfo {author} {\bibfnamefont {A.}~\bibnamefont {Hamadeh}},
  \bibinfo {author} {\bibfnamefont {F.}~\bibnamefont {Abreu~Araujo}}, \bibinfo
  {author} {\bibfnamefont {A.~D.}\ \bibnamefont {Belanovsky}}, \bibinfo
  {author} {\bibfnamefont {P.~N.}\ \bibnamefont {Skirdkov}}, \bibinfo {author}
  {\bibfnamefont {R.}~\bibnamefont {Lebrun}}, \bibinfo {author} {\bibfnamefont
  {V.~V.}\ \bibnamefont {Naletov}}, \bibinfo {author} {\bibfnamefont
  {K.}~\bibnamefont {Zvezdin}}, \bibinfo {author} {\bibfnamefont
  {M.}~\bibnamefont {M}}, \bibinfo {author} {\bibfnamefont {J.}~\bibnamefont
  {Grollier}}, \bibinfo {author} {\bibfnamefont {O.}~\bibnamefont {Klein}},
  \bibinfo {author} {\bibfnamefont {V.}~\bibnamefont {Cros}}, \ and\ \bibinfo
  {author} {\bibfnamefont {G.}~\bibnamefont {de~Loubens}},\ }\href@noop {}
  {\bibfield  {journal} {\bibinfo  {journal} {arXiv:1506.03603v1}\ } (\bibinfo
  {year} {2015})}\BibitemShut {NoStop}%
\bibitem [{\citenamefont {Colpa}(1978)}]{Colpa1978}%
  \BibitemOpen
  \bibfield  {author} {\bibinfo {author} {\bibfnamefont {J.}~\bibnamefont
  {Colpa}},\ }\href@noop {} {\bibfield  {journal} {\bibinfo  {journal} {Physica
  A: Statistical Mechanics and its Applications}\ }\textbf {\bibinfo {volume}
  {93}},\ \bibinfo {pages} {327 } (\bibinfo {year} {1978})}\BibitemShut
  {NoStop}%
\end{thebibliography}

\clearpage
\begin{center}\section*{Supplementary material}\end{center}

\section{Hamiltonian formalism}

The magnetization dynamics can be described by means of the Landau-Lifshitz equation
\begin{equation}
\label{eq:LL}
  \frac{d\vec{M}}{dt} = -\gamma\vec{M}\times \vec{H}_{eff},
\end{equation}
where $\vec{M}$ is the magnetization vector, $\gamma$ is the gyromagnetic ratio and $H_{eff}$ is an effective field. In the Hamiltonian formalism proposed by Slavin and Tiberkevich~\cite{Slavin2009}, Eq.~(\ref{eq:LL}) is recast as a function of the complex amplitude $a$ defined through a Holstein-Primakoff transformation
\begin{equation}
\label{eq:HP}
  a = \frac{m_1+im_2}{\sqrt{2M_S(M_S+m_3)}},
\end{equation}
where $m_3$ is the magnetization component parallel to $\vec{H}_{eff}$, $m_1$ and $m_2$ are perpendicular to $\vec{H}_{eff}$, and $M_S=||(m_1,m_2,m_3)||$ is the saturation magnetization density.

By expanding the resulting equation in Taylor series, the linear dynamics can be written as
\begin{equation}
\label{eq:linears}
    \frac{da}{dt} = -i\frac{d}{da^*}\mathcal{H}(a,a^*),
\end{equation}
where $\mathcal{H}$ is the Hamiltonian of the system. In the main text, we generalize Eq.~(\ref{eq:linears}) to an array of complex amplitudes $\underline{a}$, so that $\mathcal{H}$ becomes a matrix.

\section{Hamiltonian matrices}

Here, we outline the expressions for the Hamiltonian matrices for the field contributions specified in the main text.

\subsection{Anisotropy field}

We assume that the anisotropy field in the magnetic elements is dominated by shape; this is certainly the case for
the Permalloy islands that are commonly used. The demagnetizing factors in thin films are defined as $N$, $M$, and $L$ in the $1$, $2$, and $3$ directions, respectively [see Eq.~(\ref{eq:HP})]. Computing the energy functional for every island leads to the diagonal Hamiltonian matrices
\begin{eqnarray}
\label{eq:Han11}
  \mathcal{H}_{an}^{(1,1)} &=& \frac{\gamma M_S(N-M)}{2}\mathcal{I}, \\
\label{eq:Han21}
  \mathcal{H}_{an}^{(1,2)} &=& \frac{\gamma M_S[4L-2(M+N)]}{4}\mathcal{I},
\end{eqnarray}
where $\mathcal{I}$ is the $4\times 4$ identity matrix.

\subsection{External field}

The external field is considered to be homogeneous throughout the spin ice structure, with magnitude $|\vec{H}|=H_o$ and an arbitrary direction in space. To second order in $\underline{a}$, the Hamiltonian takes a diagonal form with terms proportional to the stable magnetization direction of each island. In other words, only fields parallel to each magnetic element easy axis will affect linear spin waves. Since we consider thin films, only $H_x=\vec{H}\hat{x}$ and $H_y=\vec{H}\hat{y}$ survive, and we are left with the matrices
\begin{eqnarray}
\label{eq:Hext11}
  \mathcal{H}_{ext}^{(1,1)} &=& \mathcal{O}, \\
\label{eq:Hext21}
  \mathcal{H}_{ext}^{(1,2)} &=& -2\begin{pmatrix}
	                                H_y & 0 & 0 & 0 \\
									0 & H_x & 0 & 0 \\
									0 & 0 & -H_y & 0 \\
									0 & 0 & 0 & -H_x \\
	                               \end{pmatrix},
\end{eqnarray}
where $\mathcal{O}$ is the $4\times 4$ zero matrix.

\subsection{Dipolar field}

The derivation for the dipolar field is outlined in the main text. The Hamiltonian matrices obtain for the ground state are
\begin{eqnarray}
\label{eq:Hd11s}
  \mathcal{H}_{d}^{(1,1)} &=&   \begin{pmatrix}
	                                S_x^{11} & S_c^{12} & S_x^{13} & S_c^{14} \\
									S_c^{21} & S_y^{22} & S_c^{23} & S_y^{24} \\
									S_x^{31} & S_c^{23} & S_x^{33} & S_c^{34} \\
									S_c^{41} & S_y^{24} & S_c^{34} & S_y^{44} \\
	                            \end{pmatrix} - \hat{S}_z\\
\label{eq:Hd21s}
  \mathcal{H}_{d}^{(1,2)} &=&   \mathcal{D}+\begin{pmatrix}
	                                        0 & S_c^{12} & S_x^{13} & S_c^{14} \\
											S_c^{21} & 0 & S_c^{23} & S_y^{24} \\
											S_x^{31} & S_c^{32} & 0 & S_c^{34} \\
											S_c^{41} & S_y^{42} & S_c^{43} & 0 \\
	                                        \end{pmatrix}+ \hat{S}_z.
\end{eqnarray}

The elements of the above Hamiltonian matrices contain contributions between a particular island and every other island in the structure but itself. Consequently, we can divide them in two terms: $S_{\beta,\tau\tau'}$ containing them sum between island $j$ at cell $\tau$ and every island in cell $\tau'$; and the elements $S_{\beta,jk}^{jk}$ containing the component-wise products between the islands in cell $\tau$. Using the notation where $\vec{R}_{jk,\tau\tau'}$ is the distance between island $j$ in cell $\tau$ and island $k$ in cell $\tau'$ and $\vec{R}_{jk}$ is the distance between islands $j$ and $k$ in island $\tau$, these terms along the Cartesian direction are defined as
\begin{subequations}
\begin{eqnarray}
\label{eq:appSxtt}
  S_{\hat{x},\tau\tau'} &=& \sum_{k,\tau'\neq\tau}\left[\frac{3(\vec{R}_{jk,\tau\tau'}\cdot\hat{x})^2}{\vec{R}_{jk,\tau\tau'}^5}-\frac{1}{\vec{R}_{jk,\tau\tau'}^3}\right]e^{-i\vec{q}\vec{R}_{jk}},\\
\label{eq:appSytt}
  S_{\hat{y},\tau\tau'} &=& \sum_{k,\tau'\neq\tau}\left[\frac{3(\vec{R}_{jk,\tau\tau'}\cdot\hat{y})^2}{\vec{R}_{jk,\tau\tau'}^5}-\frac{1}{\vec{R}_{jk,\tau\tau'}^3}\right]e^{-i\vec{q}\vec{R}_{jk}},\\
\label{eq:appSztt}
  S_{\hat{z},\tau\tau'} &=& \sum_{k,\tau'\neq\tau}\left[-\frac{1}{\vec{R}_{jk,\tau\tau'}^3}\right]e^{-i\vec{q}\vec{R}_{jk}},\\
\label{eq:appSxjk}
  S_{\hat{x},jk}^{jk} &=& \left[\frac{3(\vec{R}_{jk}\cdot\hat{x})^2}{\vec{R}_{jk}^5}-\frac{1}{\vec{R}_{jk}^3}\right],\\
\label{eq:appSyjk}
  S_{\hat{y},jk}^{jk} &=& \left[\frac{3(\vec{R}_{jk}\cdot\hat{y})^2}{\vec{R}_{jk}^5}-\frac{1}{\vec{R}_{jk}^3}\right],\\
\label{eq:appSzjk}
  S_{\hat{z},jk}^{jk} &=& \left[-\frac{1}{\vec{R}_{jk}^3}\right].
\end{eqnarray}
\end{subequations}

The summation on the cross direction $\hat{x},\hat{y}$ can be similarly divided into two contributions, defined as
\begin{eqnarray}
\label{eq:appSbtt}
  S_{c,\tau\tau'} &=& \sum_{k,\tau'\neq\tau}\left[\frac{3(\vec{R}_{jk,\tau\tau'}\cdot\hat{x})(\vec{R}_{jk,\tau\tau'}\cdot\hat{y})}{\vec{R}_{jk,\tau\tau'}^5}\right]e^{-i\vec{q}\vec{R}_{jk}},\\
\label{eq:appSbjk}
  S_{c,jk}^{jk} &=& \left[\frac{3(\vec{R}_{jk}\cdot\hat{x})(\vec{R}_{jk}\cdot\hat{y})}{\vec{R}_{jk}^5}\right].
\end{eqnarray}

Finally, the diagonal matrix $\mathcal{D}$ in Eq.~(\ref{eq:Hd21s}) can be labeled from $1$ to $4$, taking the values
\begin{subequations}
\begin{eqnarray}
\label{eq:D1}
  D_1 &=& S_{\hat{x},\tau\tau'}+2\left(S_{\hat{y},13}+S_{c,14}-S_{c,12}\right),\\
\label{eq:D2}
  D_2 &=& S_{\hat{y},\tau\tau'}+2\left(S_{\hat{x},24}+S_{c,23}-S_{c,21}\right),\\
\label{eq:D3}
  D_3 &=& S_{\hat{x},\tau\tau'}+2\left(S_{\hat{y},31}+S_{c,32}-S_{c,34}\right),\\
\label{eq:D4}
  D_4 &=& S_{\hat{y},\tau\tau'}+2\left(S_{\hat{x},42}+S_{c,43}-S_{c,41}\right).
\end{eqnarray}
\end{subequations}

In the case of the remanent state, we note that the components of the Hamiltonian are $4\times 4$ matrices with a similar form as the matrices in the vortex state Hamiltonian. In fact, taking the first two rows and columns of the above matrices and considering the structure's translation vector for a remanent state, $\vec{R}_{r}=(d\hat{x},d\hat{y})$, leads to the correct Hamiltonian matrices.

\section{Exchange interaction}

Intra-island exchange interactions between non-collinear spins are important to correctly describe the dynamics of spin ices, especially the modes that arise because of internal degrees of freedom. For our analytical model, we consider each island to be composed of $\mathcal{N}_{ex}$ macrospins interacting with their nearest neighbors by using an effective, discrete Heisenberg Hamiltonian model. The exchange Hamiltonian matrix blocks defined above will now have a dimension $(8\mathcal{N}_{ex})\times(8\mathcal{N}_{ex})$ to take into account the internal spins in each island. Consequently, the elements of the Hamiltonian matrices above must be also expanded by replacing each of them by an $\mathcal{N}_{ex}\times \mathcal{N}_{ex}$ block.

Furthermore, we can introduce an arbitrary direction for each spin, so that the complex amplitude of a particular spin is
\begin{eqnarray}
\label{eq:agen}
  a &=& \left[\left(1-2|a|^2\right)\cos{\theta}+\sqrt{1-|a|^2}\left(a+a^*\right)|\sin{\theta}|\right]\hat{x}\nonumber\\
  &+&\left[\left(1-2|a|^2\right)\sin{\theta}+\sqrt{1-|a|^2}\left(a+a^*\right)|\cos{\theta}|\right]\hat{y}\nonumber\\
  &-& i\sqrt{1-|a|^2}\left(a-a^*\right)\hat{z},
\end{eqnarray}
where $\theta$ is the angle with respect to the $\hat{x}$ axis, and the absolute values represent the isotropic nature of deviations from the magnetic elements' easy axis.

We consider $\mathcal{N}_{ex}=3$, i.e., one bulk and two edge spins. For the particular example of the vortex square ice, the exchange Hamiltonian takes the form
\begin{eqnarray}
\label{eq:Hex11}
  \mathcal{H}_{ex}^{(1,1)} &=& \mathcal{O}|_{12\times 12}, \\
\label{eq:Hex21}
  \mathcal{H}_{ex}^{(1,2)} &=& \mathcal{I}\otimes C,
\end{eqnarray}
where $\mathcal{O}|_{12\times 12}$ indicates a $12\times 12$ zero matrix, and $C$ is expressed as a function of the exchange constant $J$ and the array of spin angles $\underline{\theta}$. For example, for the case where all spins are collinear in a single magnetic element, the matrix $C$ takes the form
\begin{equation}
\label{eq:C}
    C = -2JM_S^2\begin{pmatrix}
                    1 & -1 & 0 \\
                    -1 & 2 & -1 \\
                    0 & -1 & 1 \\
                \end{pmatrix}
\end{equation}

The dominant Hamiltonian matrices described above can be extended simply by completing the diagonal terms in $\mathcal{H}_{an}$ and $\mathcal{H}_{ext}$ and calculating the summations between the new spins of \emph{different} islands in $\mathcal{H}_{d}$. Non-collinear spins can be also easily included in the model by computing the products originating from the definition of Eq.~(\ref{eq:agen}).

\section{Angle dependence}

\begin{figure}[t] 
\centering \includegraphics[width=5.in]{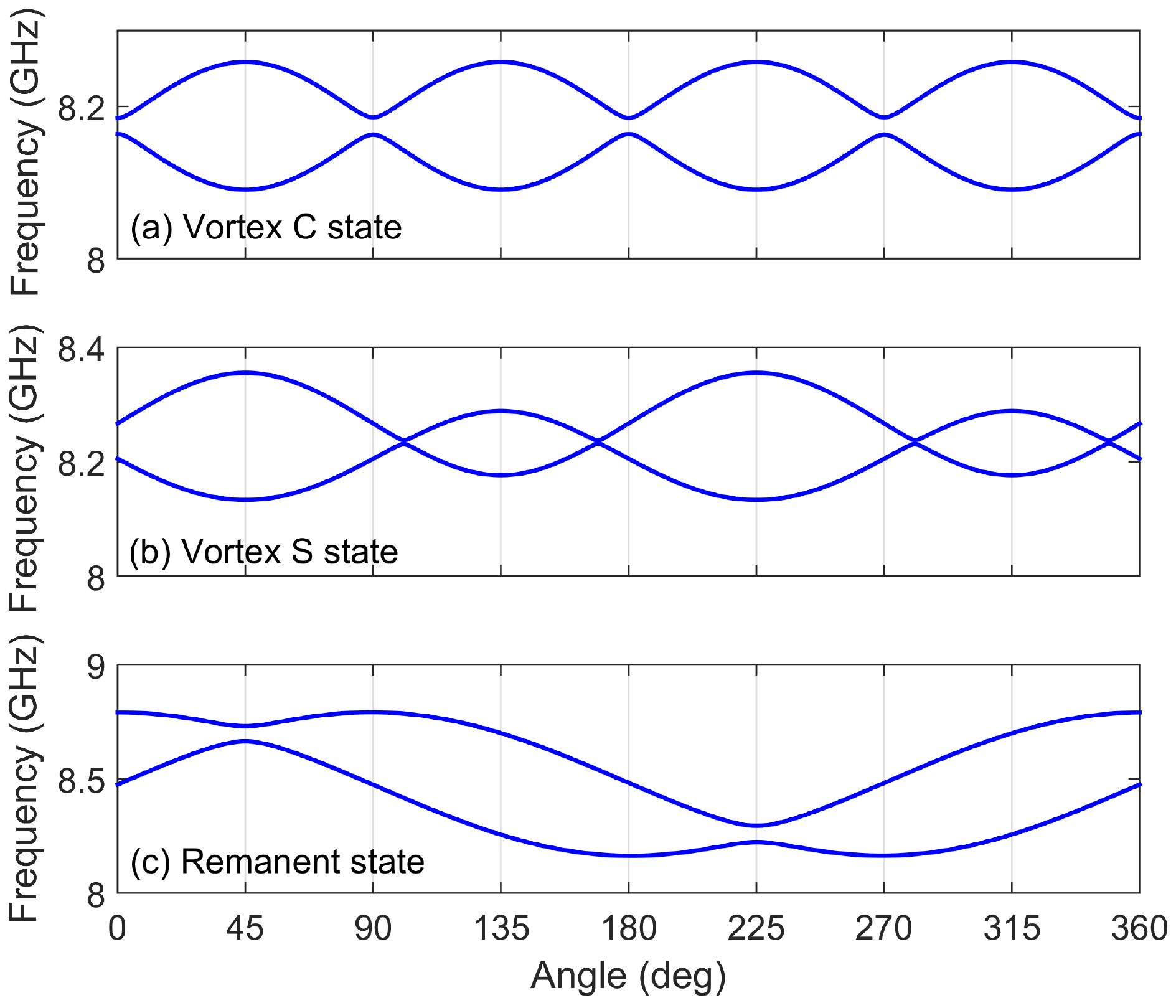}
\caption{ \label{fig1s} (Color online) Angle dependence of spin wave bands using an applied field of magnitude $50$~Oe. Panels (a) and (b) shows both the parallel modes for a C and S vortex state, exhibiting their different symmetries. The remanent state is shown in panel (c), also consistent with their symmetry. }
\end{figure}

In the main text, we explored the effect of an applied field along the $\hat{x}$ direction. Varying the angle of such a field provides the means to explore the symmetry of the square ices. As discussed in the main text, C and S states are energetically stable in the nanoislands. A vortex state with C-state magnetic elements has a four-fold symmetry while both vortex and remanent states with S-state magnetic elements have a two-fold symmetry. This can be readily shown by calculating the angle dependence of the spin waves at the $\Gamma$ point with an applied field of $50$~Oe. Figure~\ref{fig1s}(a)-(b) clearly displays these symmetries for each case, focusing on the M2 and M3 bulk modes for clarity. On the other hand, the remanent state in an onion state or in a macrospin approximation has a two-fold symmetry with elements magnetized at 90 degrees. The resulting angle dependence shown in Fig.~\ref{fig1s}(c) follows these symmetries as well. Such an angle dependence represents a valuable tool to experimentally manipulate the magnon spectra, and to infer the magnetic configuration of square ices. Coupled with the field tunability, it is possible to unambiguously determine the dominant static state throughout the structure.

\section{Micromagnetic simulations}

As specified in the main text, full-scale micromagnetic simulations implemented a computational system containing eight islands. The system was discretized into a mesh of size 1.25~nm$\times1.25$~nm$\times5$~nm and the magnetostatic interactions calculated using fast Fourier transform with periodic boundary conditions applied in the plane. The system was first set in an approximate local equilibrium state with each island homogeneously magnetized along their easy axis, approximating the vortex or remanent states. These initial configurations were then relaxed by integrating the Landau-Lifshitz-Gilbert equation for the micromagnetic spins using a dimensionless damping of $\alpha=0.25$. The magnetization of the islands in a vortex (remanent) states then relaxed into an onion (S) state.

The relaxed configuration was then subjected to a uniform external field pulse of magnitude approximately 10~Oe in the $(-\hat{x},-\hat{y})$ direction for 50~ps and the full Landau-Lifshitz-Gilbert equation integrated in time steps of 0.25~ps for 10~ns using a damping of $\alpha=0.02$. Magnetization configurations were sampled every 25~ps; the average magnetization at each time slice was Fourier transformed to yield 1D spectra of the magnetization components as functions of frequency, and the sequence of 2D time slices was Fourier transformed to yield full 2D amplitude and phase maps for each frequency. These calculations were performed for constant external fields of 0, 50, and 100~Oe along the (1,0) direction for the vortex and remanent states. 

\end{document}